%% file: BDT_paper.tex
\documentclass[superscriptaddress,
prx,
twocolumn,
preprintnumbers,
amsmath,
amssymb,
altaffilletter,
floatfix,
]{revtex4-2}

\usepackage{graphicx}
\usepackage{dcolumn}
\usepackage{bm}
\usepackage{caption}
\usepackage{subcaption}

\DeclareMathOperator*{\argmin}{arg\,min}

\def\mjd{\textsc{Majorana Demonstrator}}

\def\ge{$^{76}$Ge}

\def\znu{$0\nu\beta\beta$}
\def\vbb{$0\nu\beta\beta$}

\def\qbb{$Q_{\beta\beta}$}

\def\th{$^{228}$Th}
\def\ap{alpha}
\def\mbdt{MSBDT}
\def\abdt{$\alpha$BDT}

\begin{document}

\preprint{APS/123-QED}

\title{Interpretable Boosted Decision Tree Analysis for the {\sc Majorana Demonstrator}}



\include{authors}

\date{\today}

\begin{abstract}
The \mjd~is a leading experiment searching for neutrinoless double-beta decay with high purity germanium detectors~(HPGe). Machine learning provides a new way to maximize the amount of information provided by these detectors, but the data-driven nature makes it less interpretable compared to traditional analysis. An interpretability study reveals the machine's decision-making logic, allowing us to learn from the machine to feedback to the traditional analysis. In this work, we have presented the first machine learning analysis of the data from the \mjd; this is also the first interpretable machine learning analysis of any germanium detector experiment. Two gradient boosted decision tree models are trained to learn from the data, and a game-theory-based model interpretability study is conducted to understand the origin of the classification power. By learning from data, this analysis recognizes the correlations among reconstruction parameters to further enhance the background rejection performance. By learning from the machine, this analysis reveals the importance of new background categories to reciprocally benefit the standard \textsc{Majorana} analysis. This model is highly compatible with next-generation germanium detector experiments like LEGEND since it can be simultaneously trained on a large number of detectors.
\end{abstract}

\maketitle


\section{Introduction}\label{sec:introduction}
Neutrinoless double beta decay~(\vbb)~\cite{0vbb_review,review2019,review_2016} is a hypothetical lepton number violating process~($\Delta L=2$) beyond the standard model. The observation of \vbb~ would prove that the neutrino is its own antiparticle, also known as the Majorana particle. This is a key ingredient for leptogenesis~\cite{leptogenesis}, which is one model that explains the observed matter-antimatter asymmetry in our universe. Measuring \vbb~is a challenging task since it occurs with an ultra-long half-life ($> 10^{26}$ \,yrs)~\cite{klz_newresult,GERDAresult}. This limits the number of signal events we can collect, and requires us to reliably discover them among a plethora of backgrounds. To maximize their discovery potential, germanium-based \vbb~searches seek to operate in the quasi-background-free regime, where less than one background event is expected in the region of interest over the full lifetime of the experiment. Therefore, the ability to suppress background as much as possible is pivotal to \vbb~search experiment.

Traditional background suppression techniques are typically derived from physical first principles, which are used to define event-level reconstruction parameters. A cut is then placed upon the reconstruction parameters to minimize backgrounds while retaining signals. Because a traditional analysis begins with first-principles, interpretability is inherently built in to this approach. However, there are weaknesses with this approach as well. The actual response of a detector to a particular background source is often clouded by complex effects inherent to the detector technology that are difficult to model, reducing the effectiveness of any background rejection cuts. Furthermore, many physical effects that produce backgrounds must be handled individually, increasing the chances that a particular source of background will be neglected. Finally, unknown detector physics could also produce potential bias in reconstruction parameters, harming the performance of traditional background cuts.

Machine learning presents an alternative to the traditional first-principles approach to background rejection, and has already been proven quite successful for neutrino physics experiments~\cite{kamnet, kamcnn,KamRNN,next_cnn,microboone_cnn,nova_cnn,sno_nn,GERDAml,review_ml,project8ml}. Unlike traditional analyses, the background suppression power of machine learning algorithms comes primarily from data. This allows machine learning models to efficiently handle unknown backgrounds to reach state-of-the-art performance. Unfortunately, learning from data makes machine learning analyses less interpretable compared to the traditional ones. Therefore, many machine learning analyses are equipped with an interpretability study to reveal the underlying decision-making logics~\cite{interp_physics1,interp_physics2,interp_physics3}.

In this work, we present the first machine learning analysis for the \mjd~\cite{MJDPRL,MJD_AHEP,MJD_PRC}, which is also the first interpretable machine learning analysis of any germanium detector experiment. This analysis was inspired by the drift-time correction to our multi-site and surface \ap~discrimination parameters, which indicated that accounting for correlations between parameters could enhance background suppression power. We constructed two boosted decision tree~(BDT) models to reject two of the most critical backgrounds in the \mjd, namely the \mbdt~for multi-site events and the \abdt~for \ap~events. Both models take individual reconstruction parameters as inputs and are trained on the detector data to provide background suppression. By learning from the data, this model utilizes multivariate correlations among reconstruction parameters to improve the background suppression. It also reduces the need for detector- and run-level tuning, which would be time-consuming in future large-scale experiments such as LEGEND~\cite{legend_pcdr}.

In addition, we conducted a comprehensive interpretability study to understand the source of classification power. This study leverages a coalitional game theory concept to unravel the black-box that is the inside of a machine learning model~\cite{SHAP}. It has been widely used in biomedical science~\cite{SHAP_bio,SHAP_lifescience,SHAP_covid} and other fields~\cite{,SHAP_pharmacy,Shap_material,SHAP_education}. By learning from the machine, we verified our BDTs' abilities to learn multivariate correlations among different features. Furthermore, we revealed the importance of new background categories that the traditional, first-principles-based analysis did not address, which eventually led to new analysis cuts in the standard \textsc{Majorana} analysis.

The paper is structured as follows. Section~\ref{sec:mjd_experiment} describes the \mjd~experiment, the major background sources and the traditional analysis cuts to reject them. Section~\ref{sec:data_pipeline} describes the data pipeline for collecting and pre-processing the training data. Section~\ref{sec:bdt} describes the gradient BDT algorithms. Section~\ref{sec:result} reports the training results of the \mbdt~and the \abdt~with a comparison to the standard \textsc{Majorana} analysis. Section~\ref{sec:interpretability} describes the interpretability study we conducted. We highlight Section~\ref{subsec:interp_abdt}, which outlines the ability of machine learning to reveal the importance of new background categories and reciprocally benefit the standard \textsc{Majorana} analysis pipeline.


\section{\sc Majorana Demonstrator}\label{sec:mjd_experiment}

The \mjd~experiment searches for \vbb~decay in \ge~ using 40.4\,kg of high purity germanium (HPGe) detectors~\cite{MJDPRL}. Of these, 27.2\,kg of p-type point-contact~(PPC) HPGe detectors are enriched to 88\% in $^{76}$Ge~\cite{processing}. The \textsc{Demonstrator} is operated at the 4850-ft level of the Sanford Underground Research Facility in Lead, South Dakota. Data were taken from August 2015 to March 2021, and are split into 9 data set (DS) periods, referred to as DS0-DS8. Starting with DS8 (August 2020), novel p-type inverted-coaxial point-contact (ICPC) detectors~\cite{ICPC} were added to the \mjd~detector array. Data taking finished in March 2021, with a total enriched exposure of 64.5\,kg$\cdot$yr, 2.82\,kg$\cdot$yr of which is from ICPC detectors~\cite{Majorana_prl}. The \textsc{Demonstrator}’s HPGe detectors, in combination with low-noise electronics~\cite{electronic_paper}, have achieved good linearity over a broad energy range~\cite{adc_nonlinearity}, and best-in-field energy resolution with a full-width-at-half-maximum (FWHM) approaching 0.1\% at the \qbb~(2039 keV) of $^{76}$Ge~\cite{MJD_PRC}. This excellent energy performance, coupled with the low energy threshold and low-background of the \textsc{Demonstrator}, makes it a competitive \vbb~decay experiment.
\begin{figure}[h]
    \centering
    \includegraphics[width=1.0\linewidth,,trim={1pc 1pc 1pc 1pc},clip]{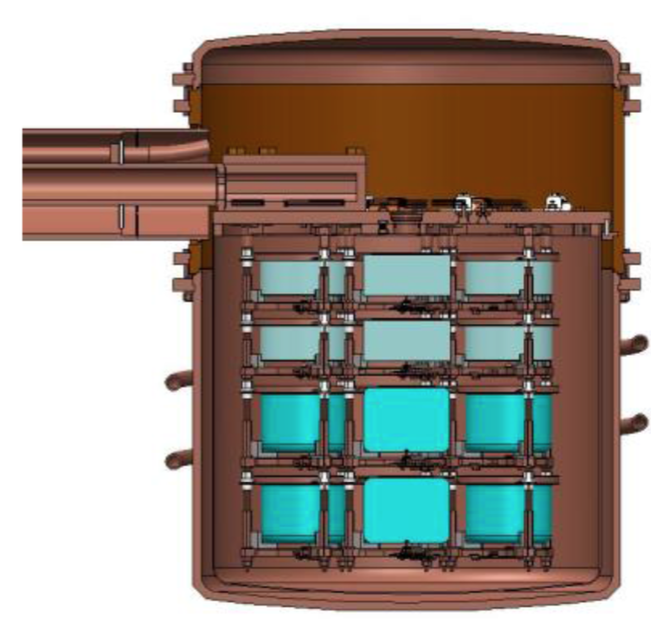}
    \caption{A diagram of one \mjd~detector module and the HPGe detectors within.}
    \label{fig:mjd_schematic}
\end{figure}

A weekly calibration is conducted to monitor detector stability and provide data for developing analysis cuts. The thorium isotope \th~was selected as the primary calibration source because its decay chain emits several gamma-rays spanning from a few hundred to 2615\,keV, which covers the \qbb~of $^{76}$Ge and allows for calibration over a wide energy range. During calibrations, the \th~source is deployed into the calibration track, which surrounds the cryostat in a helical path~\cite{mjd_calibration}. Event energies are tuned to minimize \th~calibration source gamma line width~\cite{charge_trapping}. This routine calibration provides an excellent source of training data that will be discussed in Section~\ref{sec:data_pipeline}.

\begin{figure*}[ht!]
    \begin{subfigure}{0.50\linewidth}
      \includegraphics[width=1.0\linewidth,trim={2.5pc 0pc 0pc 0pc},clip]{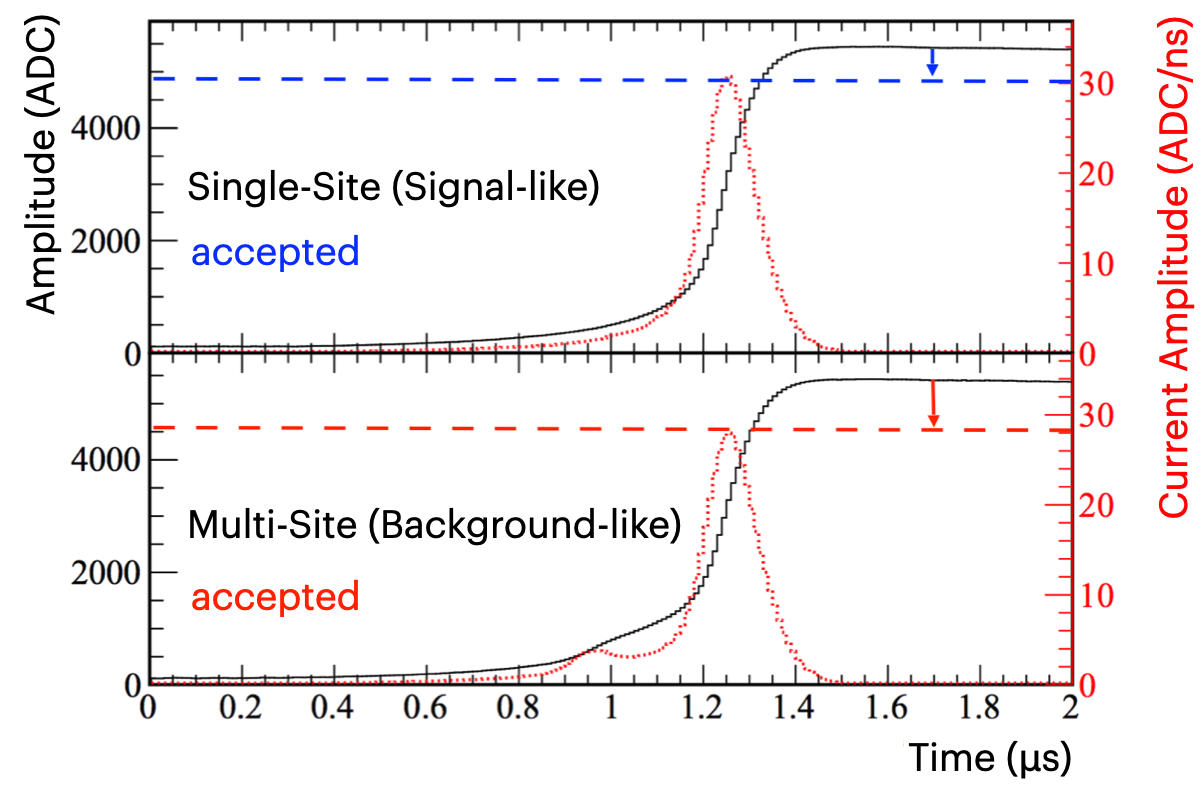}
      \caption{}
      \label{fig:ss-ms}
    \end{subfigure}
    \begin{subfigure}{0.47\linewidth}
      \includegraphics[width=1.0\linewidth,trim={0pc 0pc 0pc 0pc},clip]{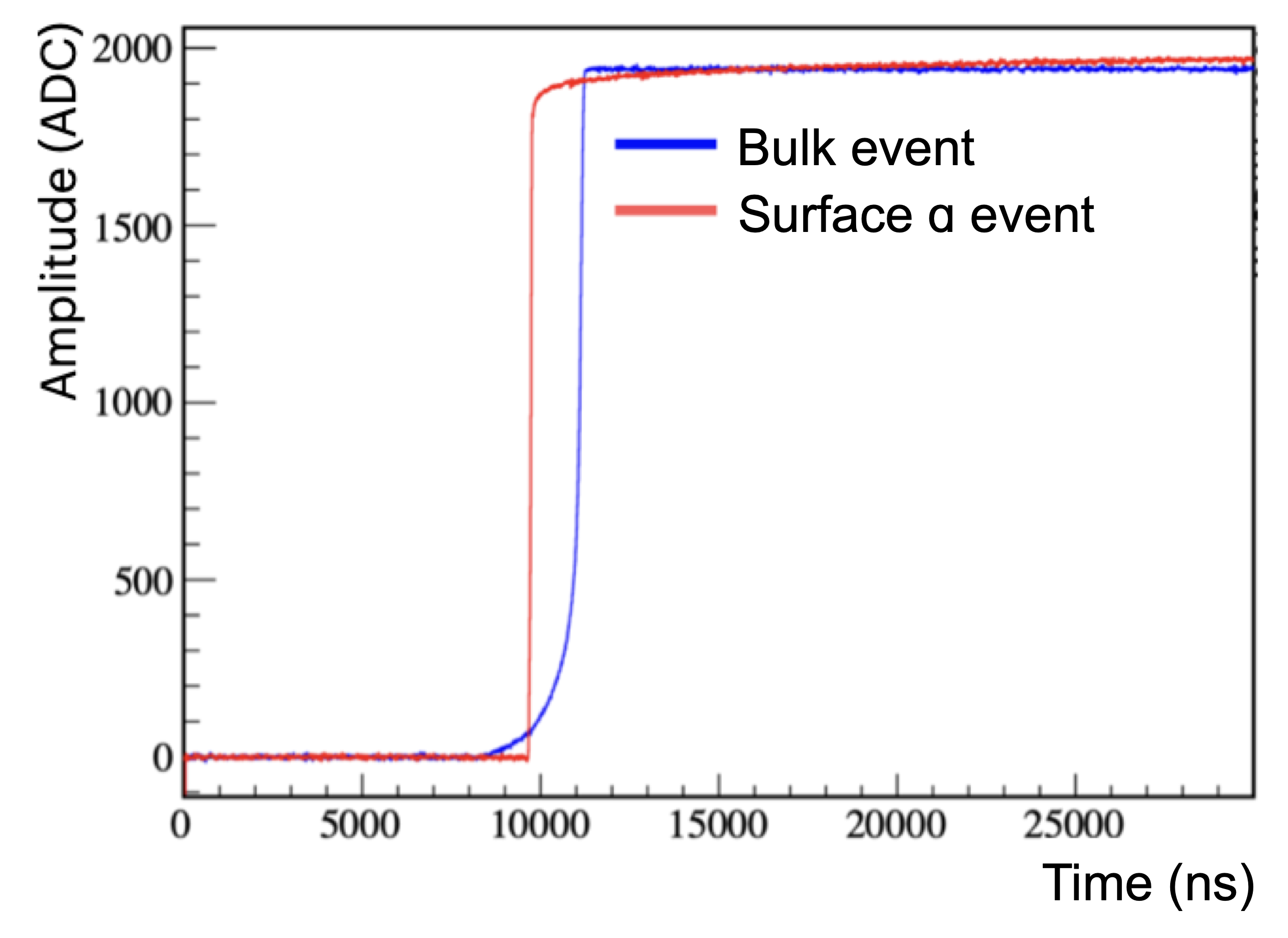}
      \caption{}
      \label{fig:alpha_ev}
    \end{subfigure}
    \caption{(a) Pulse shape plot of single-site events (top) and multi-site events (bottom). The black line shows the raw waveform in ADC counts and the red line shows the waveform in current amplitude. (b) Illustration of waveforms from surface \ap~event and bulk event. }
    \label{fig:mjd_waveforms}
\end{figure*}

Most \vbb~events are single-site events which deposit all of their charge in a single location of  $<$1 mm linear dimension in detector. This type of event appears in a \mjd~detector as a waveform with a single sharply-rising step, as indicated by the black trace in the top panel of Figure~\ref{fig:ss-ms}. Since the waveform itself is the integrated ionized charge collected from an energy deposition in the detector, the derivative of the waveform (red traces in the same panel) is effectively the current induced as charges drift towards the point contact. In the \mjd, two major background sources are multi-site events and surface-\ap~events. If charge is deposited at multiple locations within the crystal, the drift times may differ up to $\sim$1\,$\mu$s, resulting in a waveform with multiple steps as shown in the bottom panel of Figure~\ref{fig:ss-ms}. This leads to a current pulse with a smaller maximum value than that of a single-site event with the same energy. Based on this first principle, we designed the current amplitude vs. energy~(\texttt{AvsE}) described in Reference~\cite{avse_paper}. The current amplitude is estimated by a linear fit to a smaller range of the waveform. Cutting on the energy-normalized current amplitude (or \texttt{AvsE}) leads to efficient multi-site event rejection. In the standard \textsc{Majorana} analysis, we select events from a dedicated \texttt{AvsE} range by applying both low and high \texttt{AvsE} cuts.

The other major source of backgrounds at \qbb~ is from the \ap~particles impacting the passivated surface and p+ contact surfaces of detectors. Prior to the most recent \mjd~data release~\cite{Majorana_prl}, this background source was rejected entirely using the first principle of the ``delayed charge recovery'' effect~\cite{TUBE_paper}. Based on the characteristics of \ap~interactions, it appears that charge mobility is drastically reduced on or near the passivated surface. Therefore, a fraction of the charge from these interactions is slowly released on the timescale of waveform digitization, leading to a measurable increase in the slope of the waveform tail. The delayed charge recovery effect lowers the peak amplitude and as shown in Figure~\ref{fig:alpha_ev} results in a slowly rising tail slope that distinguishes this event from a bulk event at the same energy. The delayed charge recovery~(\texttt{DCR}) cut tags events with larger tail slopes to efficiently reject surface \ap s.

At a later stage of the standard \textsc{Majorana} analysis, a novel analysis cut based on the first principles of late charge~(LQ) was developed. The \texttt{LQ} cut probes the top of the rising edge of the waveform to identify delayed charge collection on $~\sim 1\mu s$ timescale. It efficiently eliminates events in the transition layer and an additional population of near-point-contact events.

The final \textsc{Majorana Demonstrator} standard analysis is described in Reference~\cite{Majorana_prl}. The standard \textsc{Majorana} analysis for the PPC detectors is developed with the Germanium Analysis Toolkit~(GAT), and is thus referred to as the ``GAT analysis". The standard \textsc{Majorana} analysis for the ICPC detectors is developed independently from GAT, and is referred to as the ``ORNL analysis". Both the GAT and ORNL analyses contain independently developed pulse shape discrimination~(PSD) cuts derived from the first principles of HPGe detector charge collection: current amplitude versus energy, the delayed charge recovery effect, as well as the late charge effect. In this paper, we denote those cuts as GAT \texttt{AvsE}/\texttt{DCR}/\texttt{LQ} cut and ORNL \texttt{AvsE}/\texttt{DCR}/\texttt{LQ} cut, respectively.

During the development of the standard \textsc{Majorana} analysis, we observed that the PSD parameters vary with the length of time it takes the charges to drift to the p+ electrode, defined as drift-time, due to well-understood charge cloud diffusion and bulk charge trapping effects. Therefore, a drift-time correction was made to correct for this correlation. In the following text, we will use the terms ``standard \texttt{AvsE}/\texttt{DCR}/\texttt{LQ}" to refer to the GAT analysis for PPC detectors and ORNL analysis for ICPC detectors, respectively. Additionally, we extracted the raw \texttt{AvsE} and raw \texttt{DCR} parameters, which are preliminary versions of standard \texttt{AvsE}/\texttt{DCR} thereby not directly used by the standard \textsc{Majorana} analysis. The raw parameters are generated under the GAT framework with a detector- and run-wise energy calibration applied. However, the raw parameters are not drift-time corrected, thus underperforming the standard \textsc{Majorana} analysis parameters. In this work, we decided to use the raw \texttt{AvsE}/\texttt{DCR} parameters to train the BDTs, and then compare the training results to the standard \textsc{Majorana} analysis parameters. The \texttt{LQ} parameters are introduced at a later stage of the standard \textsc{Majorana} analysis, thus we decided not to incorporate it to train the machine learning analysis. However, we did include \texttt{LQ} when comparing the two analyses in Section~\ref{subsec:analysis_compare}. 
\section{Data Pipeline}\label{sec:data_pipeline}
We collected both a signal and a background dataset to train the BDT. The signal dataset should be representative of the signal (\znu~events in our case), and the background dataset should represent the proper background to reject. Before creating signal and background datasets from the \textsc{Majorana} data, a standard suite of cuts is applied: periods of high noise associated with liquid nitrogen fills or unstable operation are removed; non-physical waveforms, pileup waveforms, and pulser events are then removed by data cleaning cuts; and finally events in which multiple germanium detectors are triggered are removed. We particularly avoided the usage of high-level selection cuts, such as standard \texttt{AvsE}/\texttt{DCR}/\texttt{LQ}, since the tuning and validation of these cuts can be time-consuming. Decoupling from these cuts allows a fast-track application of this model on newly-taken data from multiple detectors. We then chose events in the double escape peak (DEP) from \th~calibration data as the signal dataset. The DEP events are pair production events where both gammas have successfully escaped from the detector, thus mostly single-site events. An energy cut of $1592.5\pm2.5$\,keV is applied to select DEP events. Monte Carlo simulations including X-ray excitations and bremsstrahlung predict the events under DEP selection criteria to be 90\% single-site with 10\% multi-site impurities.
\begin{center}
\renewcommand{\arraystretch}{1.0}
\begin{table}[hb!]
\centering
\caption{List of input features to the \mbdt~and the \abdt. Cat.\ stands for categorical and Cont.\ stands for continuous.} 
\begin{tabular}{c|c|c}
\hline
Features & Type & Description \\
\hline
\texttt{detType} & Cat.\ & \begin{tabular}{@{}c@{}} detector type: \\ enriched PPC or ICPC\end{tabular}\\
\hline
\texttt{channel} & Cat.\ & detector DAQ channel\\
\hline
\texttt{tDrift} & Cont. &  \begin{tabular}{@{}c@{}} drift time from the start of the rise\\ to  99\% waveform amplitude\end{tabular}\\
\hline
\texttt{tDrift50} & Cont. &  \begin{tabular}{@{}c@{}} drift time from the start of the rise\\ to 50\% waveform amplitude\end{tabular}\\
\hline
\texttt{AvsE} & Cont. &\begin{tabular}{@{}c@{}} raw A. vs E, pulse shape parameter\\ for multi-site event rejection\cite{avse_paper,Majorana_prl}\end{tabular} \\
\hline
\texttt{DCR} & Cont. & \begin{tabular}{@{}c@{}} raw DCR, pulse shape parameter\\ for \ap~event rejection~\cite{TUBE_paper,Majorana_prl}\end{tabular} \\
\hline
\texttt{noise} & Cont. &\begin{tabular}{@{}c@{}}  measuring 10-20MHz noise\end{tabular}\\
\hline
\texttt{DS} & Cat.\ & \begin{tabular}{@{}c@{}}  data period the event belongs to,\\ defined by run ranges\end{tabular}\\
\hline
\end{tabular}
\label{tab:param_list}
\end{table}
\end{center}
The background datasets for the \mbdt~and the \abdt~are selected separately. For MSBDT, we select events under the single escape peak~(SEP) of \th~calibration data. The SEP events are pair production events where only one gamma has escaped from the detector. Although all SEP events are technically multi-site, if those sites occur at the isochrones of equal drift-time, they will reach the point contact at roughly the same time. In that case, the time difference between the two sites is smaller than the detector's timing resolution, resulting in an apparently single-site waveform even though more than one energy deposition has occurred. These events are the impurities to the background dataset. An energy cut of $2103.5 \pm 2.5$\,keV is applied to select these events.

For the \abdt, since the major \ap~backgrounds are energy-degraded alpha events from the continuum, we have to select training data from a broader spectrum. We first select high energy alpha events by collecting events above 2615\,keV in background runs. This sample is expected to have some contamination from high energy gamma events, originating from the decay of cosmogenically- and neutron-induced isotopes and uranium/thorium chain. We then select low energy alpha events by collecting the \texttt{DCR} tagged background events in a 1000-2615\,keV energy range. In this way, a total of 723 high energy alpha and 2,839 low energy alpha events are selected. 

After event selection, we extract 8 features from every event. The names and descriptions of these parameters are listed in Table~\ref{tab:param_list}. \texttt{AvsE} and \texttt{DCR} are dedicated pulse shape parameters for multi-site/alpha rejection respectively. Other parameters are added to probe their multivariate correlations with \texttt{AvsE}/DCR and to each other. For example, adding the \texttt{channel} parameter will allow the model to perform detector-wise tuning; adding \texttt{tDrift} and \texttt{tDrift50} will allow the model to perform a drift-time correction, and our \texttt{noise} parameter allows us to look for correlations during noisy periods in the data. Among all features in Table~\ref{tab:param_list}, some features are continuous and some features are categorical. BDTs naturally handle both types of feature in the structure of the tree, allowing us to train on all detectors from all run periods simultaneously.

\subsection{Data Augmentation}\label{subsec:data_aug}
The data we selected above are highly imbalanced. First of all, only about 4\% of data points are provided by ICPC detectors while the rest are from PPC detectors. Secondly, only 3,562 $\alpha$ events are collected compared to ~600,000 DEP events in \abdt~training. This forms a typical long tail distribution where the head class contains most of the events and the tail class contains only a minimal proportion. If a BDT is trained with such an imbalanced dataset, it will be heavily biased towards the ample head class while ignoring the scarce tail class. We fix this issue by performing data augmentation. 

Data augmentation refers to algorithms that generate synthetic data points to boost the population of the tail class for training purposes. We employ it to boost the population of both ICPC detector events and surface \ap~events. The input dataset contains 8 features per event, 3 of which are categorical features. A Synthetic Minority Over-sampling TEchnique - Nominal and Continuous~(SMOTE-NC)~\cite{smote} algorithm is adopted for data augmentation. SMOTE-NC generates synthesized data by randomly interpolating between datapoints and its nearest neighbors. It works well on low dimensional data with both continuous and categorical features. It is first applied to all 3 datasets~(DEP, SEP and \ap) to boost the population of ICPC events by a factor of $\sim$50, then applied again on the \ap~dataset to boost the population of \ap~events by a factor of $\sim$115. We refer to the events directly collected from detectors as genuine events and the events from data augmentation as augmented events. The augmented events are only used for training; model evaluation that will be discussed in Section~\ref{sec:result} is based on genuine events.
\subsection{Distribution Matching}
While building our BDT models, we want them to look at the correlations of features instead of single features, unless that single feature is the first-principle feature as discussed in Section~\ref{sec:mjd_experiment}. The first-principle feature---that is \texttt{AvsE} for MSBDT or \texttt{DCR} for $\alpha$BDT---is designed to fulfill the same background rejection goal as the BDT model. We do not expect features other than the first-principle feature to contribute to the classification independently, but they can contribute through their correlations with the first-principle feature or other features. Undesirable behavior arises when other parameters are allowed to contribute independently to classification. For example, if a given \texttt{channel} in the \mbdt~training dataset is accidentally biased to contain 50\% more signals than backgrounds, the BDT will ``remember" this bias and tend to classify events in this \texttt{channel} as signal regardless of the rest of the features. If we then validate the BDT on another out-of-sample, unbiased dataset, the classification performance on this \texttt{channel} will be suppressed. This phenomenon is referred to as overfitting.
\begin{figure}[htb!]
    \centering
    \includegraphics[width=0.95\linewidth,trim={0.1pc 0.5pc 3pc 2.8pc},clip]{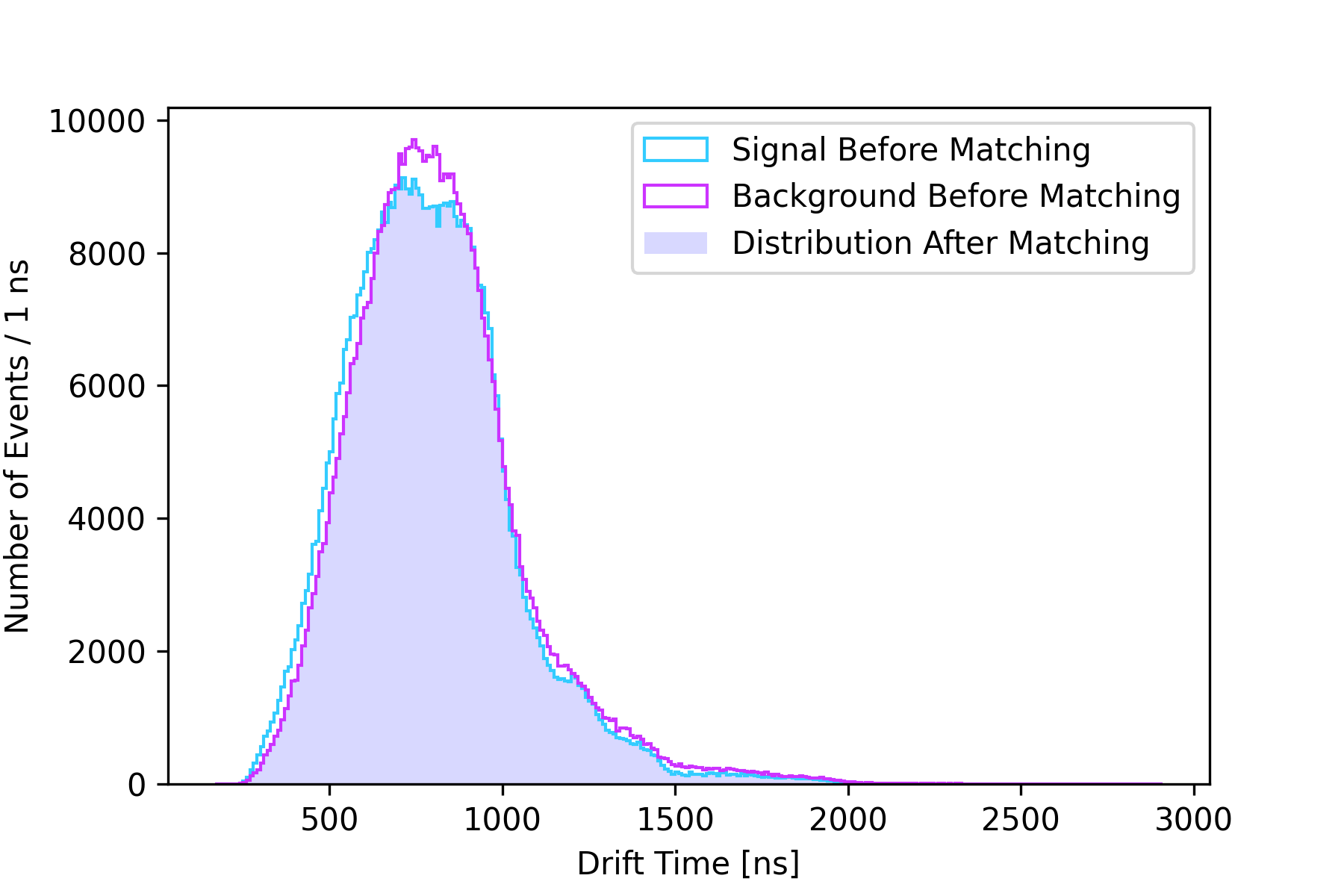}
    \caption{Distribution matching of the \texttt{tDrift} feature in input data.}
    \label{fig:dist_match}
\end{figure}
To avoid this kind of overfitting, we carried out a process called ``distribution matching" on 6 out of 8 secondary features. Figure~\ref{fig:dist_match} shows the distribution matching effect on the \texttt{tDrift} feature. The distribution of each secondary feature is first put into a histogram with predefined bin width. Then for every bin in the histogram, we randomly sample without replacement the same number of events from the signal and background datasets. This will reduce the size of both signal and background datasets to the same amount. Eventually, the sampled events are aggregated into a new signal/background dataset. We leave the first-principle feature---AvsE for MSBDT and DCR for \abdt---unmatched because we expect them to follow different distributions between the signal and background dataset. The \texttt{detType} feature is not matched either since it overlaps with the \texttt{channel} feature. After distribution matching, the signal/background dataset spectrum will exhibit the same spectrum shape over matched secondary features. Distribution matching only affects the training dataset. The performance of trained BDT is evaluated on both DEP/SEP datasets and a flat Compton Continuum(CC) dataset, as described in Section~\ref{subsec:analysis_compare} and Table~\ref{tab:rejection_efficiency}.

In summary, the data pipeline for the \mbdt~contains the following steps: we first select \th~DEP events as the signal dataset and \th~SEP events as the background dataset; we then extract eight features as described in Table~\ref{tab:param_list} for every event in both datasets; we then perform data augmentation to generate augmented ICPC events; lastly, we perform distribution matching on all features except \texttt{AvsE} and \texttt{detType}. The data pipeline of the \abdt~ contains the following steps: we first select \th~DEP events as the signal dataset, and aggregate both low- and high-energy \ap s to form the genuine alpha dataset; we then extract eight features; perform data augmentation to generate augmented ICPC detector events and \ap~events as the background dataset; lastly, we perform distribution matching on all features except \texttt{DCR} and \texttt{detType}.

\section{Boosted Decision Tree}\label{sec:bdt}
The decision tree~(DT) model produces classification decisions by making a series of binary choices. This features allow decision tree to naturally handle both continuous and categorical dataset, without the need of additional structures such as one-hot encoding. Boosting algorithms allow the machine to generate many decision trees iteratively to form a classification ``committee". After training the m$^{\mathrm{th}}$ decision tree, the classification committee containing the first through m$^{\mathrm{th}}$ trees is denoted $T_{m}(x_{i})$. The dataset can be described as $\{x_{i},y_{i}\}_{i=0}^{k}$, where $x_{i}$ is the input event, $y_{i}$ is the label and k is the number of events. The dataset is modified according to the output of $T_{m}(x_{i})$. The modified dataset is then fed into $T_{m+1}(x_{i})$ for training. The way the dataset is modified for each iteration defines the boosting algorithm type. In this work, the BDT model is trained using the LightGBM package~\cite{lightgbm}. LightGBM adopts a gradient boosting algorithm~\cite{gradient_boost} to grow decision trees. First, a binary cross-entropy loss function $L(y_{i}, T_{m}(x_{i}))$ is defined for the classification task, where $y_{i}$ is the event label. Then for each data point, we calculate the pseudo-residual $r_{im}$:
\begin{equation}
    r_{im} = -\frac{\partial L(y_{i}, T_{m}(x_{i}))}{\partial T_{m}(x_{i})}
\end{equation}
$r_{im}$ is the negative gradient of the loss function with respect to the classification committee output at $x_{i}$. For each boosting iteration, the dataset is modified from $\{x_{i},y_{i}\}_{i=0}^{k}$ to $\{x_{i},r_{im}\}_{i=0}^{k}$, then a new decision tree $h_{m+1}(x)$ is fit to the modified dataset. The new decision tree is incorporated into the committee via the following equation:
\begin{equation}
    T_{m+1}(x) = T_{m}(x) + \gamma_{m+1}h_{m+1}(x)
\end{equation}
Where $\gamma_{m+1}$ is chosen to minimize the loss function by solving the following optimization problem:
\begin{equation}
    \gamma_{m+1} = \argmin_{\gamma}\sum_{i=0}^{k}L(y_{i},T_{m}(x_{i}) + \gamma h_{m+1}(x_{i}))
\end{equation}

The procedure above describes the mathematical formulation of BDT training. To train the BDT model, we first mix and shuffle the signal and background datasets. We then split the mixed dataset into training and validation datasets with an 80:20 ratio. The BDT models are trained on the training dataset. An early stopping algorithm will terminate the training process if the loss on the validation dataset does not decrease for a given number of iterations. The performance is quantified on a dedicated evaluation dataset which will be discussed later in Section~\ref{sec:result}; the interpretability study is conducted on a customized interpretability dataset, which will be discussed later in Section~\ref{sec:interpretability}.

LightGBM contains several highly customizable BDT models defined by collections of hyperparameters. Hyperparameters refer to parameters that do not change during training, such as the type of boosting algorithms, maximum number of trees to grow, number of early stopping iterations, and maximum number of leaves per tree. Some hyperparameters may greatly impact the metric, while other parameters may have minimal to no impact. All hyperparameters are searched simultaneously using Bayesian optimization to maximize the background rejection efficiencies at 90\% signal acceptance~\cite{botorch}.

\section{Result}
\label{sec:result}
\begin{figure*}[hbt!]
    \begin{subfigure}{0.47\linewidth}
      \includegraphics[width=1.0\linewidth,trim={0pc 0pc 0pc 0pc},clip]{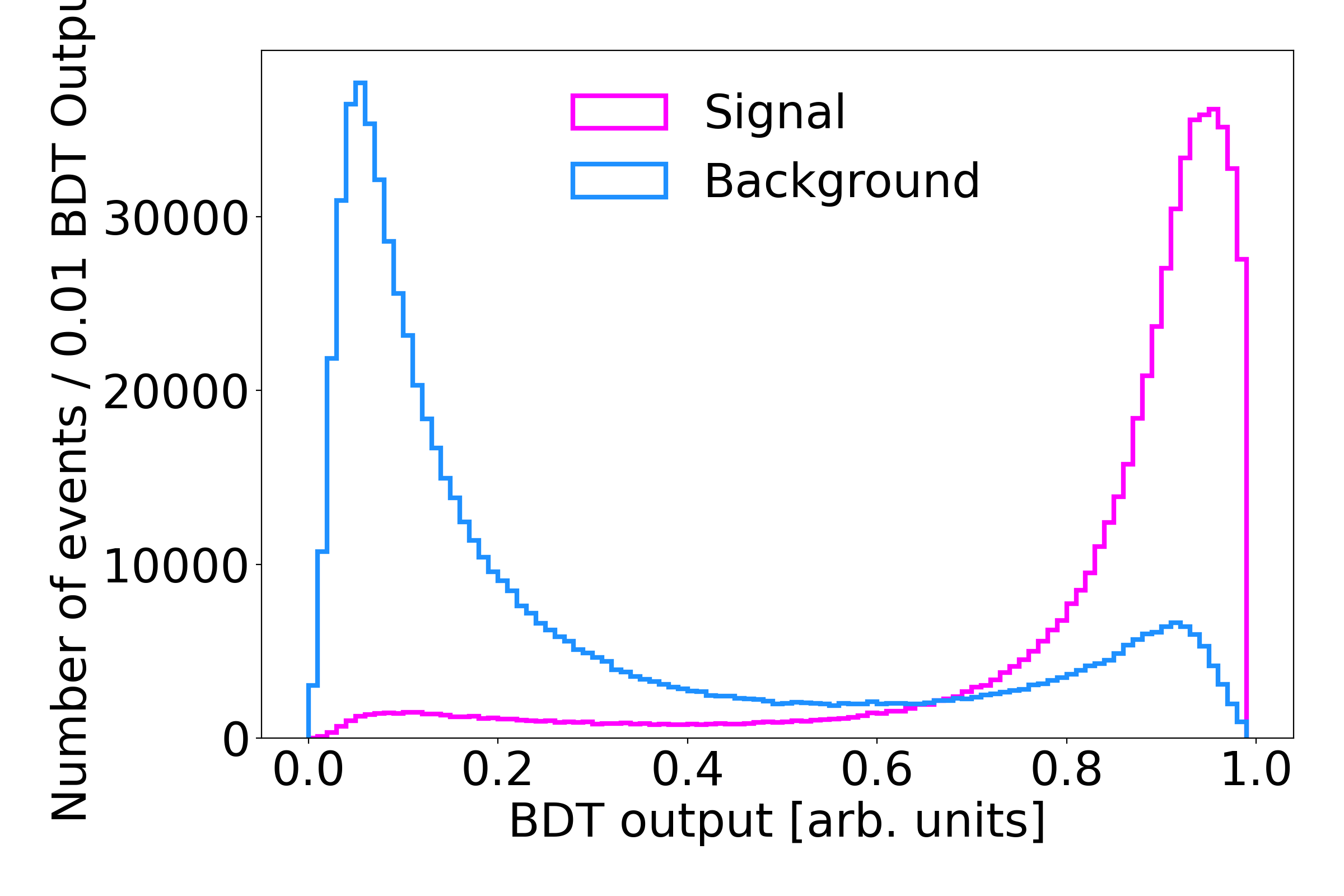}
      \caption{}
      \label{fig:MSBDT_output}
    \end{subfigure}
    \begin{subfigure}{0.51\linewidth}
      \includegraphics[width=1.0\linewidth,trim={0pc 0pc 0pc 0pc},clip]{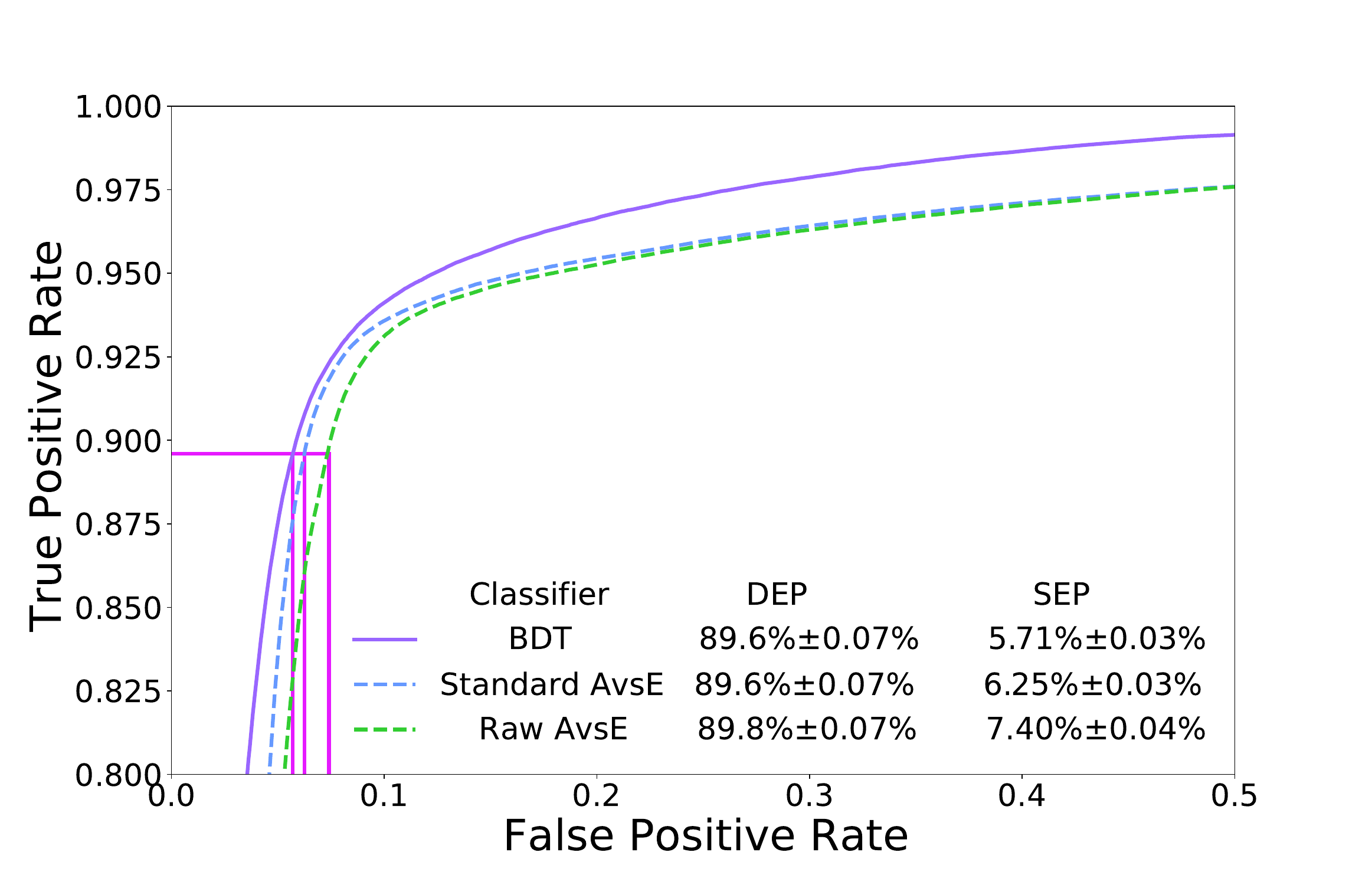}
      \caption{}
      \label{fig:MSBDT_ROC}
    \end{subfigure}
    \caption{(a) MSBDT score distribution for signal and background events. (b) Background subtracted ROC curve for MSBDT classifier, \texttt{AvsE} corrected classifier and \texttt{AvsE} classifier. The ROC curve plots the true positive rate~(TPR) vs. the false positive rate~(FPR) of a binary classifier by placing the cutting threshold at every possible location. Larger area under ROC curve represents better classification performance. For both \texttt{AvsE} classifiers, only the traditional low \texttt{AvsE} cut are applied. }
   \label{fig:pressure_map}
\end{figure*}
\begin{figure*}[hbt!]
    \begin{subfigure}{0.47\linewidth}
      \includegraphics[width=1.0\linewidth,trim={0pc 0pc 4pc 0pc},clip]{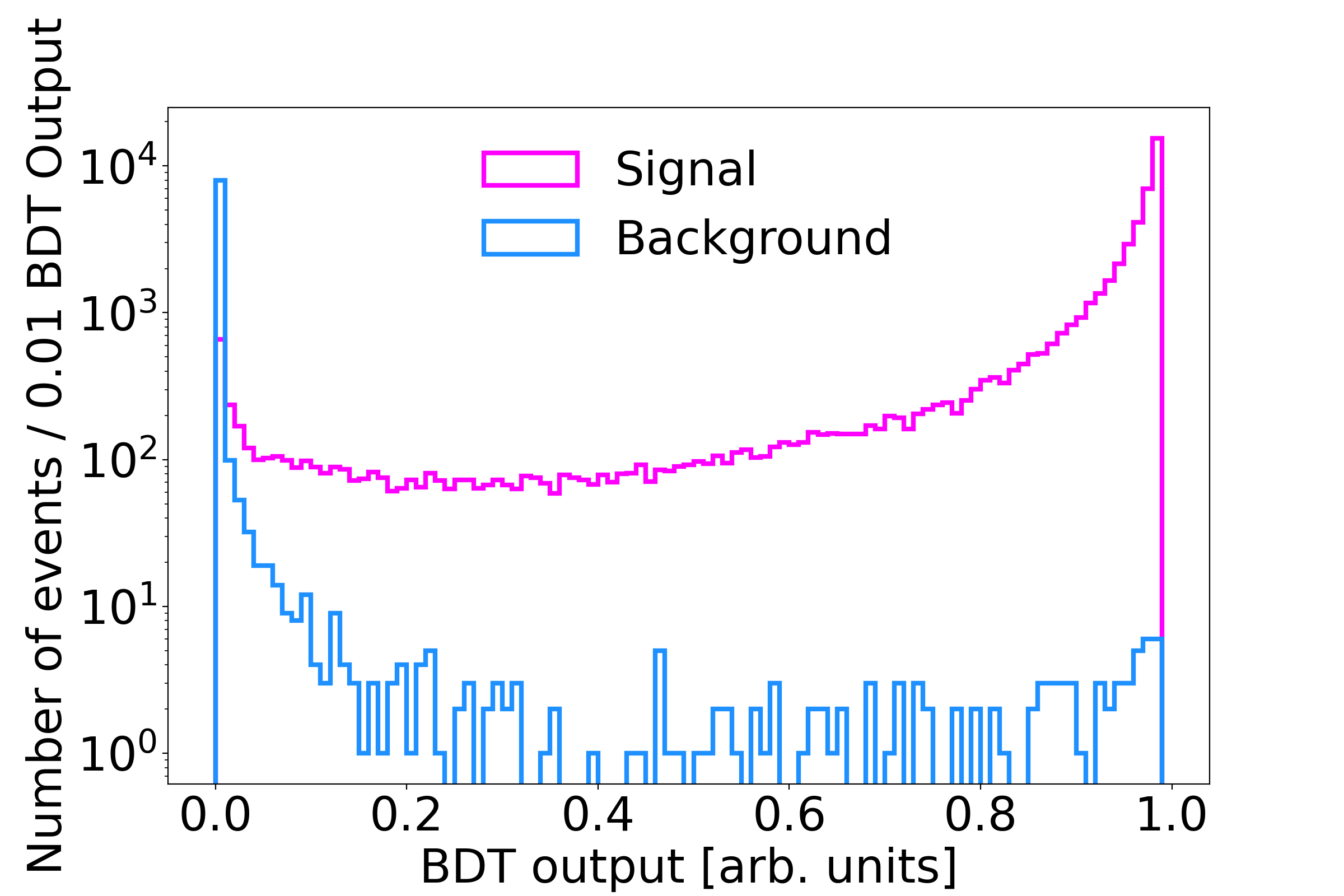}
      \caption{}
      \label{fig:aBDT_output}
    \end{subfigure}
    \begin{subfigure}{0.50\linewidth}
      \includegraphics[width=1.0\linewidth,trim={1pc 0pc 0pc 0pc},clip]{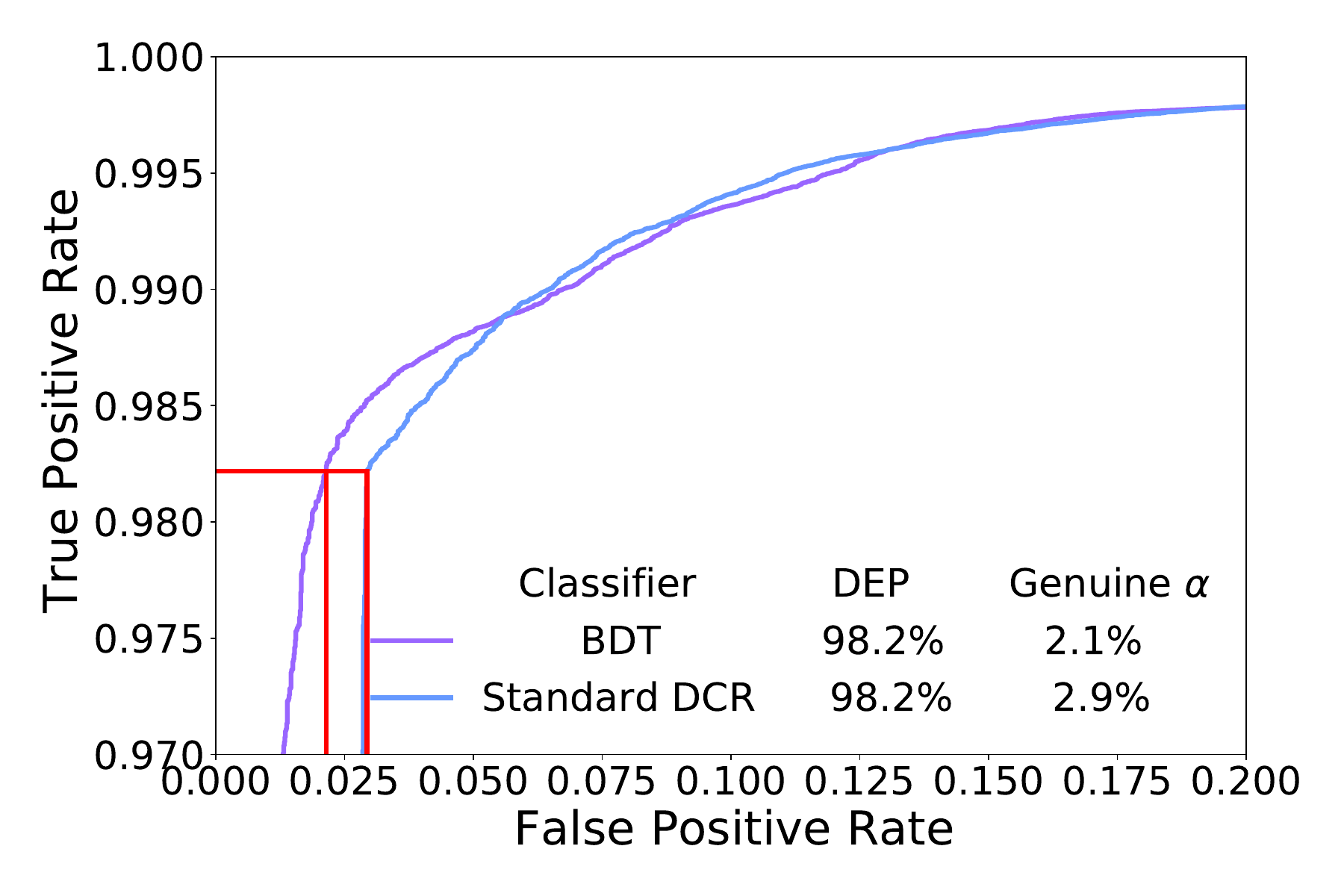}
      \caption{}
      \label{fig:aBDT_ROC}
    \end{subfigure}
    \caption{(a) \abdt~output distribution for signal and background events. (b) ROC curve for \abdt~classifier and standard \texttt{DCR} classifier.}
   \label{fig:pressure_map}
\end{figure*}

After training, we evaluate the performance of both the \mbdt~and the \abdt. The trained BDT model takes the input of 8 features as described in Table~\ref{tab:param_list}, and outputs a single floating point number as the classification score between 0 and 1. A higher classification score indicates the input event is more signal-like and a lower classification score indicates the input event is more background-like. A cutting threshold is placed to accept signals and reject backgrounds. The selection criteria of the cutting threshold will be discussed in the following subsections. We also use the Receiver Operating Characteristic~(ROC)~\cite{ROC_paper} curve to gauge the classification performance of our models. The ROC curve plots the true positive rate~(TPR) vs. the false positive rate~(FPR) by placing the cutting threshold at every possible location. The fraction of area under the ROC curve~(AUC) statistically describes the classification power of a binary classifier, in that larger AUC corresponds to better classification performance and smaller AUC corresponds to worse classification performance. For example, an AUC of 1 indicates perfect classification, and an AUC of 0.5 indicates no classification.


\subsection{MSBDT Result}\label{subsec:MSBDT_result}
Events from the \th~calibration data sets are used to test the \mbdt. The DEP events from 1590-1595 \,keV are used as the signal event sample, and the SEP events from 2101-2106\,keV are used as the background event sample. The MSBDT score spectra for signal and background samples are illustrated in Figure~\ref{fig:MSBDT_output}. The signal and background peaks are well-separated, albeit with ``misclassified'' events under both peaks. Since both the signal and the background samples have impurities described in Section~\ref{sec:data_pipeline}, these events could be the impure events and thus being correctly classified.

\begin{table*}
\renewcommand{\arraystretch}{1.5}
\caption{Table of survival fractions of signal and background events in each \textsc{Majorana Demonstrator} dataset. The signal and background event selection is defined in Section~\ref{subsec:MSBDT_result} for MSBDT and Section~\ref{subsec:aBDT_result} for the \abdt~. The BDT cutting thresholds are selected so that they produce the same signal acceptance with the standard analyses. The AvsE corrected, \texttt{DCR} corrected and \texttt{LQ} parameters are adopted in the standard \textsc{Majorana} analysis. CC stands for the compton continuum (1989-2089\,keV) events in \th~calibration dataset, and BEW stands for the \vbb~background estimation window events (1950-2350\,keV, excluding three gamma peaks) in the \vbb~search dataset. The survival numbers in BEW are calculated after joint cuts. The number in parentheses (last column of Row 11) is the survival number without the \texttt{LQ} cut.} 
\begin{tabular}{c|c|cccccccccccc|c}
\hline
Row&Dataset&DS0&DS1&DS2&DS3&DS4&DS5&DS5c&DS6&DS6c&DS7&DS8&DS8&All DS\\
Index&Detector Type&PPC&PPC&PPC&PPC&PPC&PPC&PPC&PPC&PPC&PPC&PPC&ICPC&(Expo. Weighted)\\
\hline
1&Exposure(kg $\cdot$ \textup{yr})&1.13&2.24&1.13&0.96&0.26&4.49&2.34&24.52&13.25&4.44&6.41&2.74&64.5\\
\hline
2&Single Site Signal (\%)&90.3&89.8&88.6&89.9&89.4&89.1&87.4&89.4&89.8&89.7&89.7&88.7&89.5\\
3&MSBDT Bkg. (\%)&5.62&5.85&5.01&5.71&6.31&5.95&5.70&5.73&5.58&4.57&6.41&5.76&5.71\\
4&Standard \texttt{AvsE} Bkg. (\%)&6.13&6.29&5.93&5.31&5.48&6.24&6.51&6.25&6.39&6.00&6.78&6.17&6.25\\
5&MSBDT Cal. CC (\%)&41.9&38.9&36.8&39.7&42.1&41.9&42.6&42.1&40.5&35.6&32.5&31.4&40.3\\
6&Standard \texttt{AvsE} Cal. CC (\%)&43.1&42.9&41.5&41.0&42.1&42.0&41.6&42.3&42.7&42.1&43.3&35.0&42.3\\
\hline
7&Bulk Event Signal (\%)&97.9&97.8&98.1&98.9&98.0&97.8&97.8&98.5&98.3&98.5&98.6&97.6&98.2\\
8&\abdt~Bkg. (\%)&0.4&1.4&2.1&0.8&1.2&3.8&3.4&1.6&1.9&3.5&4.7&8.1&2.1\\
9&Standard \texttt{DCR} Bkg. (\%)&1.7&1.9&2.8&3.9&0.0&2.9&5.7&2.6&3.1&5.4&2.8&0.8&2.9\\
\hline
10&BDT \vbb~BEW (\#)&11&6&2&0&0&8&6&66&23&20&19&3&164\\
11&Standard \vbb~BEW (\#)&11&4&1&0&0&9&5&58&20&17&24&4&153 (168)\\
\hline
\end{tabular}
\label{tab:rejection_efficiency}
\end{table*}

The ROC curves of MSBDT, standard \texttt{AvsE}, and raw \texttt{AvsE} are shown in Figure~\ref{fig:MSBDT_ROC}. At each cutting location, the baseline subtraction and uncertainty evaluation are then performed in the same way as in Reference~\cite{avse_paper}. To set the \mbdt~cutting threshold, we first apply the standard \texttt{AvsE} cut---that is, \texttt{AvsE} $>-1.0$---to the evaluation dataset. This cut leads to a TPR of 89.6\%, shown as the horizontal magenta line in Figure~\ref{fig:MSBDT_ROC}. Next, the BDT cutting threshold and raw \texttt{AvsE} cutting threshold are selected to reach the same TPR as the standard \texttt{AvsE}. At this level of TPR, the survival fraction of background samples of the raw \texttt{AvsE}, the standard \texttt{AvsE} and \mbdt~are 7.40\%, 6.25\%, and 5.71\%, respectively. The standard \texttt{AvsE} leverages the drift-time correlations to reject 16.3\% of SEP events that the raw \texttt{AvsE} accepts. MSBDT leverages additional multivariate correlations to reject a further 8.6\% of SEP events that the standard \texttt{AvsE} accepts.

Rows 2-4 in Table~\ref{tab:rejection_efficiency} compares the performance of the \mbdt~and the standard \texttt{AvsE} for each \textsc{Majorana} dataset. For most datasets, the \mbdt~outperforms the standard \texttt{AvsE} on selected data samples. This demonstrates the ability of our BDTs to self-discover the drift-time corrections and other possible feature correlations to improve background rejection performance. Meanwhile, introducing \texttt{DS}, \texttt{channel}, and \texttt{detType} as categorical features allows the machine to perform detector- and run-level tuning without explicitly programming. However, the background data samples described above are only good representations of the true background dataset; the deviation from true background dataset could come from energy ~(DEP energy vs. \qbb~energy) and subtle differences in the intra-detector distribution of event positions. Therefore, additional data samples are collected to examine model performance near the true energy region of interest of \vbb~decay. These data samples---denoted as Calibration Compton continuum (Cal. CC) samples, contain all events between 1989 and 2089\,keV from the \th~calibration runs. Only 40.3\% of Cal. CC samples survive \mbdt~while 42.3\% survive standard \texttt{AvsE} as shown in Row 5-6, Table~\ref{tab:rejection_efficiency}.

\subsection{$\alpha$BDT Result} \label{subsec:aBDT_result}
Events from the \th~calibration data sets and the \vbb~search data sets are used to test the \abdt. All \th~calibration events between 1000\,keV and 2380\,keV that pass the standard AvsE cut are selected as the signal samples, and the collection of genuine \ap~events are selected as the background samples. The \abdt~output distribution of signal and background datasets are shown in Figure~\ref{fig:aBDT_output}. Based on the plot, the background dataset is highly concentrated near 0.0 BDT score, indicating an excellent \ap~tagging efficiency of the \abdt. Meanwhile, the signal dataset spans the entire range, but most events are still concentrated near 1.0 \abdt~output.

The ROC curves of the \abdt~and the standard \texttt{DCR} parameter are shown in Figure ~\ref{fig:aBDT_ROC}. A cutting threshold is set at the horizontal red line to accept 98.2\% of signal events. This acceptance matches the standard \texttt{DCR} acceptance in the standard \textsc{Majorana} analysis. At this cutting threshold, the \texttt{DCR} corrected analysis has 2.9\% background acceptance, while the \abdt~only accepts 2.1\% of surface \ap~events. As mentioned in Section~\ref{sec:data_pipeline}, low energy genuine \ap~are \texttt{DCR} tagged background events between 1000\,keV and 2615\,keV. Therefore, when evaluating the performance of the standard \texttt{DCR} cut, 100\% of low energy genuine \ap~will be manifestly removed. Given the fact that standard \texttt{DCR} is ``cheating'' on low energy \ap~rejection, the \abdt~ still outperforms standard \texttt{DCR} by rejecting 27.6\% of genuine \ap~events that standard \texttt{DCR} accepts.

\subsection{Comparison to Standard {\sc Majorana} Analysis}
\label{subsec:analysis_compare}

\begin{figure*}[hbt!]
    \centering
    \includegraphics[width=0.95\linewidth,,trim={12pc 0pc 12pc 0pc},clip]{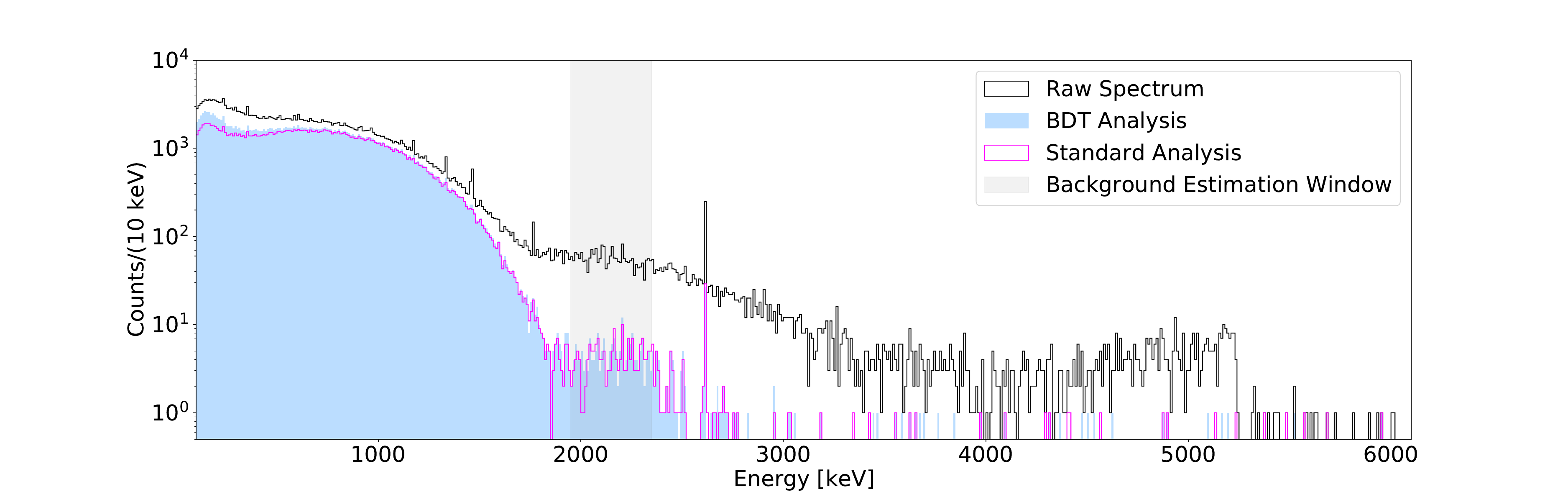}
    \caption{Energy spectrum of the standard \textsc{Majorana} analysis and the BDT analysis with 64.5$\pm$0.9\,kg exposure, a good agreement is reached within the background estimation window. The low energy discrepancy between two analyses was mainly due to the \texttt{LQ} cut.}
    \label{fig:spectrum}
\end{figure*}

The background index of the standard \textsc{Majorana} analysis is evaluated in the \vbb~Background Estimation Window~(\vbb~BEW). The \vbb~BEW samples are collected from 1950 to 2350 keV, excluding $\pm$5\,keV region around the 2039\,keV \qbb~value and three gamma peaks at 2103\,keV, 2118\,keV and 2204\,keV. Based on simulations, the background rate is expected to be flat after the exclusion. Both BDTs are applied to produce a number of survival events for \vbb~BEW in each dataset, which can be compared to the number of survival events for each dataset after applying the suites of standard \textsc{Majorana} analysis cuts: the standard \texttt{AvsE} cut, the high \texttt{AvsE} cut, the \texttt{DCR} cut, and the \texttt{LQ} cut. Note that neither the \texttt{LQ} feature nor the transition layer events are included in the BDT training process; thus, the BDT analysis will not be sensitive to these events. Given this ``unfair'' condition, the BDT analysis still manages to match and, in some datasets, outperform the standard \textsc{Majorana} analysis. The total number of \vbb~BEW survival events for BDT/standard \textsc{Majorana} analysis are 164 and 153~\cite{Majorana_prl}, indicating consistency between the two. As a comparison, standard \textsc{Majorana} analysis without \texttt{LQ} cut allows 168 events to remain in the \vbb~BEW. Figure~\ref{fig:spectrum} shows the comparison between the two analyses over the entire energy range. The two analyses agrees well except in the low energy region, where the standard \textsc{Majorana} analysis cuts more aggressively. This discrepancy is mainly subject to the the \texttt{LQ} cut, and the applicability of \texttt{LQ} at low energy is still under investigation. Therefore, these agreements show that the BDT analysis can start from raw parameters and tune them to match a highly optimized analysis.

\subsection{ICPC Detectors}\label{subsec:ICPC_result}

For the ICPC detectors, the trained BDT model takes the raw parameters as input and compares the output BDT score to the ORNL analysis parameters. This leads to additional challenges since the raw parameters are developed with GAT while the ORNL parameters are independently developed and customized for ICPC detectors. In this analysis, we use the raw parameters as inputs to train the BDT models to reach or exceed the background rejection performance of the ORNL analysis. This means that the BDT must perform multivariate corrections and account for the technical differences between two independently developed analyses. The second-to-last column of Table~\ref{tab:rejection_efficiency} shows the model evaluation results on ICPC detectors. MSBDT outperforms the ORNL AvsE for multi-site event rejection, with a background survival fraction of only 5.76\%, compared to 6.17\% for the ORNL AvsE. It also makes a significant improvement over its primary input parameter, raw AvsE, which has a 25\% background survival fraction (not shown in Table II). On the other hand, the \abdt~underperforms the ORNL DCR with a 8.1\% genuine alpha survival fraction, compared to 0.8\% for the ORNL DCR. However, the \abdt~still makes a significant improvement over its primary input parameter, raw DCR, which allows 18\% of genuine alphas to survive (not shown in Table II). The BDT analyses account for the technical differences among independently developed analyses to simultaneously analyze different types of germanium detectors. Finally, the result of the BDT analyses indicate that the GAT analysis has the potential to reach the same level of performance on ICPC detectors under proper tuning. 

\section{Machine Interpretability}\label{sec:interpretability}

We demonstrated the BDT's ability to outperform the standard \textsc{Majorana} analysis, but the source of additional classification power was not readily apparent. To identify these sources, a post facto machine interpretability study was performed on the trained \mbdt~and \abdt. This study used a coalitional game theory concept, Shapley value~\cite{shapley_val}, to interpret the decision of a BDT. The Shapley value is defined as follows:
\begin{equation}
    \phi_{i}(v) = \sum_{S\subseteq N/\{i\}}\frac{|S|!(n-|S|-1)!}{n!}(v(S\cup\{i\})-v(S))
\end{equation}
$v$ is the characteristic function that maps a subset of players to a real number. $S$ represents a coalition of players without the player $i$. $N$ is the set of all players and $n$ is the size of that set. In this analysis, $v$ is the BDT model mapping input features to the BDT score. Each feature is a ``player" of the game. $v(S\cup\{i\})-v(S)$ describes the difference in BDT score including/excluding feature $i$. This difference is summed over all coalitions S --- that is, the possible combinations of all other features except $\{i\}$ --- to produce the Shapley value for $i$. Therefore, the Shapley value in the context of a BDT represents each feature’s contribution to the final BDT score, assuming they work collaboratively.

The interpretability study was conducted using the SHAP package~\cite{SHAP}. The underlying mechanism is analogous to a one-dimensional free body diagram~\cite{force_plot}. SHAP assigns a Shapley value to each feature of the events to be interpreted. The Shapley value acts as a ``force" to change the BDT score: a positive Shapley value pushes the BDT score toward a more signal-like score, while a negative Shapley value pushes the BDT score toward a more background-like score. After all ``forces" are applied, the BDT reaches an equilibrium, and the equilibrium position is the BDT score of the input event. Therefore, if an event is classified as signal, the feature with the largest positive Shapley value will be the driving factor for this classification decision, while features with negative Shapley values suggest against the classification decision. An example force plot of a single \mjd~event is shown in Figure~\ref{fig:force_plot}. By investigating the Shapley values on designated datasets, we will understand the driving feature which leads to the additional classification power.

\subsection{Interpreting MSBDT}\label{subsec:interp_msbdt}

\begin{figure*}[hbt!]
        \begin{subfigure}{0.48\linewidth}
        \centering
        \includegraphics[width=1.0\linewidth,,trim={1pc 0pc 0pc 0.5pc},clip]{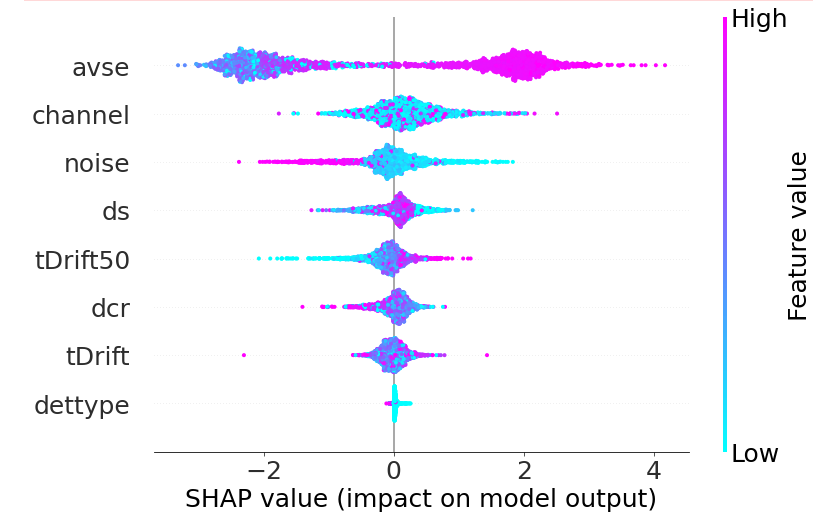}
        \caption{}
        \label{fig:msbdt_feature}
    \end{subfigure}
    \begin{subfigure}{0.48\linewidth}
      \includegraphics[width=1.0\linewidth,,trim={0 0pc 0pc 0pc},clip]{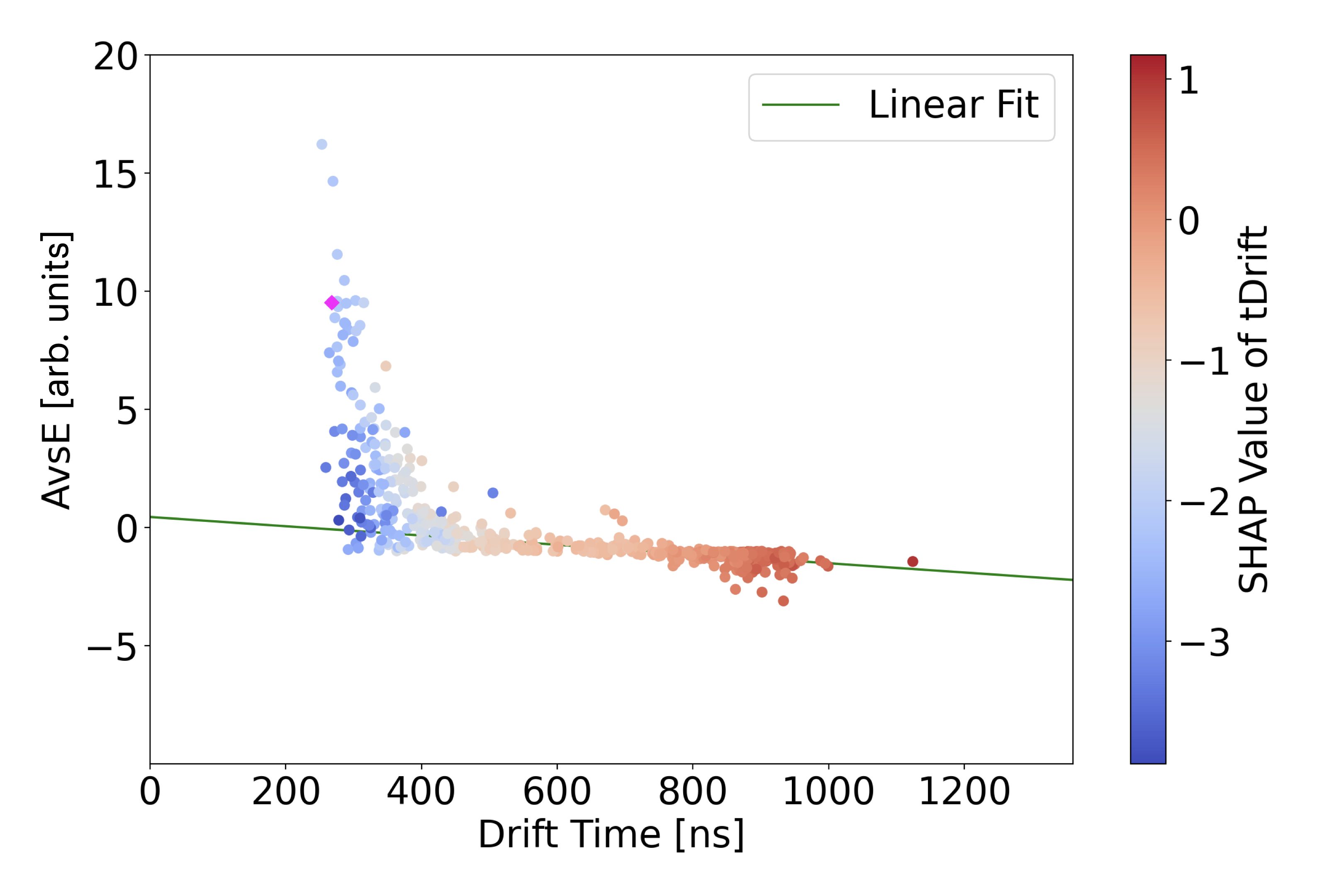}
    \caption{}
    \label{fig:tdrift_dependency}
    \end{subfigure}\\
    \begin{subfigure}{\linewidth}
      \includegraphics[width=1.0\linewidth,trim={0pc 0pc 0pc 0pc},clip]{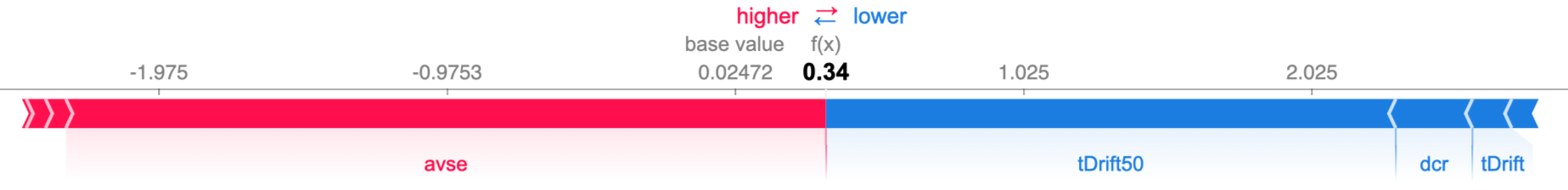}
    \caption{}
    \label{fig:force_plot}
    \end{subfigure}
    \caption{(a) Feature importance plot of the \mbdt. From top to bottom, the features are ranked from the most important to the least important. The color of each dot represents the Shapley value of each feature, normalized by the highest and lowest Shapley values of all input samples. (b) The scatter plot of raw \texttt{AvsE} vs. \texttt{tDrift} for outperforming events in a single detector in DS6. The color of each dot represents the sum of Shapley value assigned to both \texttt{tDrift} and \texttt{tDrift50}. Higher magnitude represents more important contributions from drift time. The green line represents a linear fit to the linear dependency of drift time. (c) Force plot of a single \mjd~event denoted by the magenta diamond in (b). The Shapley value of \texttt{AvsE} and \texttt{channel} provides positive forces, while the Shapley value of \texttt{tDrift}, \texttt{tDrift50}, and \texttt{dcr} provides negative forces. The equilibrium position is at 0.34. }
   \label{fig:pressure_map}
\end{figure*}

Figure~\ref{fig:msbdt_feature} presents a summary plot to illustrate the feature importance of the \mbdt. To make this plot, we first randomly sampled 10,000 \th~DEP events and 10,000 \th~SEP events to form the interpretability dataset. The Shapley values are calculated for each event in the dataset, and the distribution of Shapley values with respect to each input feature is plotted in Figure~\ref{fig:msbdt_feature}. The shape of these distributions represents the importance of the given feature. An important feature exhibits a dumbbell shape, indicating this feature drives the decision by a large magnitude most of the time. A less important feature exhibits a spindle shape, indicating this feature outputs a near-zero Shapley value most of the time but occasionally drives the decision with a large amplitude. An irrelevant feature exhibits a vertical bar shape, indicating that this feature almost always outputs a Shapley value of 0. Figure~\ref{fig:msbdt_feature} ranks the importance of features from top to bottom according to this rule. The most important feature is \texttt{AvsE} as we expected, and the second most important feature is \texttt{channel}. This means MSBDT's classification power mainly comes from channel-wise calibration of \texttt{AvsE}. The least important feature is \texttt{detType} since it is redundant with \texttt{channel}. The importance ranking shown in Figure~\ref{fig:msbdt_feature} can also be used for feature selection. In case the computation power is limited, low-importance features such as \texttt{detType} can be removed from the input. In this work, the BDT training takes less than one miunte on CPU. Therefore, low-importance features are kept since they do not seem to adversely affect the performance.

To further understand the classification power of MSBDT, especially the additional classification power compared to standard \texttt{AvsE}, we collected outperforming events from the interpretability dataset with two criteria:
\begin{itemize}
    \item DEP events that MSBDT classifies as signal but raw \texttt{AvsE} classifies as background
    \item SEP events that MSBDT classifies as background but raw \texttt{AvsE} classifies as signal
\end{itemize}
Figure~\ref{fig:tdrift_dependency} shows the joint distribution of drift time and raw \texttt{AvsE} on a 2D scatter plot on outperforming events. To avoid smearing caused by different detectors or different datasets, only outperforming events from a single detector in DS6 are shown. The color of each dot indicates the summed Shapley value of \texttt{tDrift} and \texttt{tDrift50}. Two types of drift time dependencies are observed on raw \texttt{AvsE}: a linear dependency appears on the raw \texttt{AvsE} for large drift time events, and a non-linear dependency on low drift time events. In the standard \textsc{Majorana} analysis, the linear dependence is corrected through a detector-by-detector drift time correction as discussed in Section~\ref{sec:mjd_experiment}. From the \mbdt's perspective, the BDT assigns a positive Shapley value on drift time to reproduce the drift time correction: although the linear dependency leads to lower-than-usual \texttt{AvsE} at large drift time, the \mbdt~successfully captures this dependency and produces a positive Shapley value to compensate for this effect. This is equivalent to the drift-time correction in standard \textsc{Majorana} analysis. Without explicit programming, the \mbdt~independently learns these correlations from data and leverages them to further improve background rejection performance as expected. 
\begin{figure*}[hbt!]
        \begin{subfigure}{0.48\linewidth}
        \centering
        \includegraphics[width=1.0\linewidth,,trim={1pc 0pc 0pc 0pc},clip]{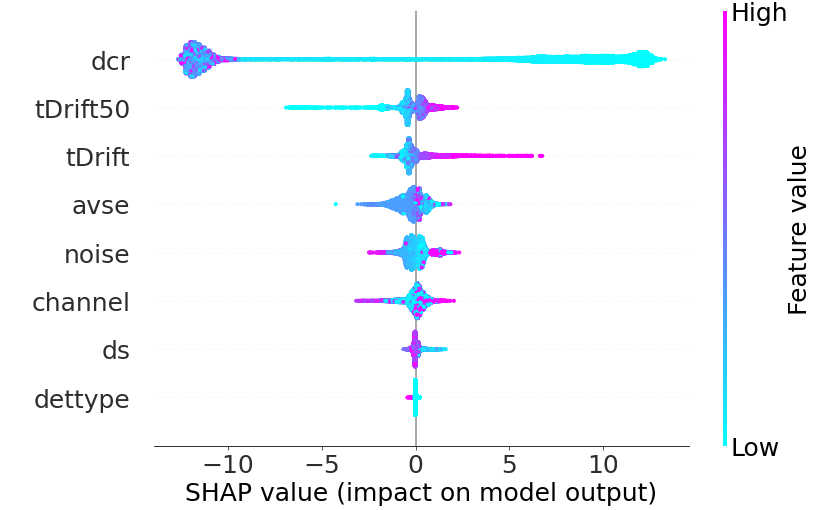}
        \caption{}
        \label{fig:abdt_feature}
    \end{subfigure}
    \begin{subfigure}{0.48\linewidth}
        \centering
        \includegraphics[width=1.0\linewidth,,trim={0 0pc 5pc 2pc},clip]{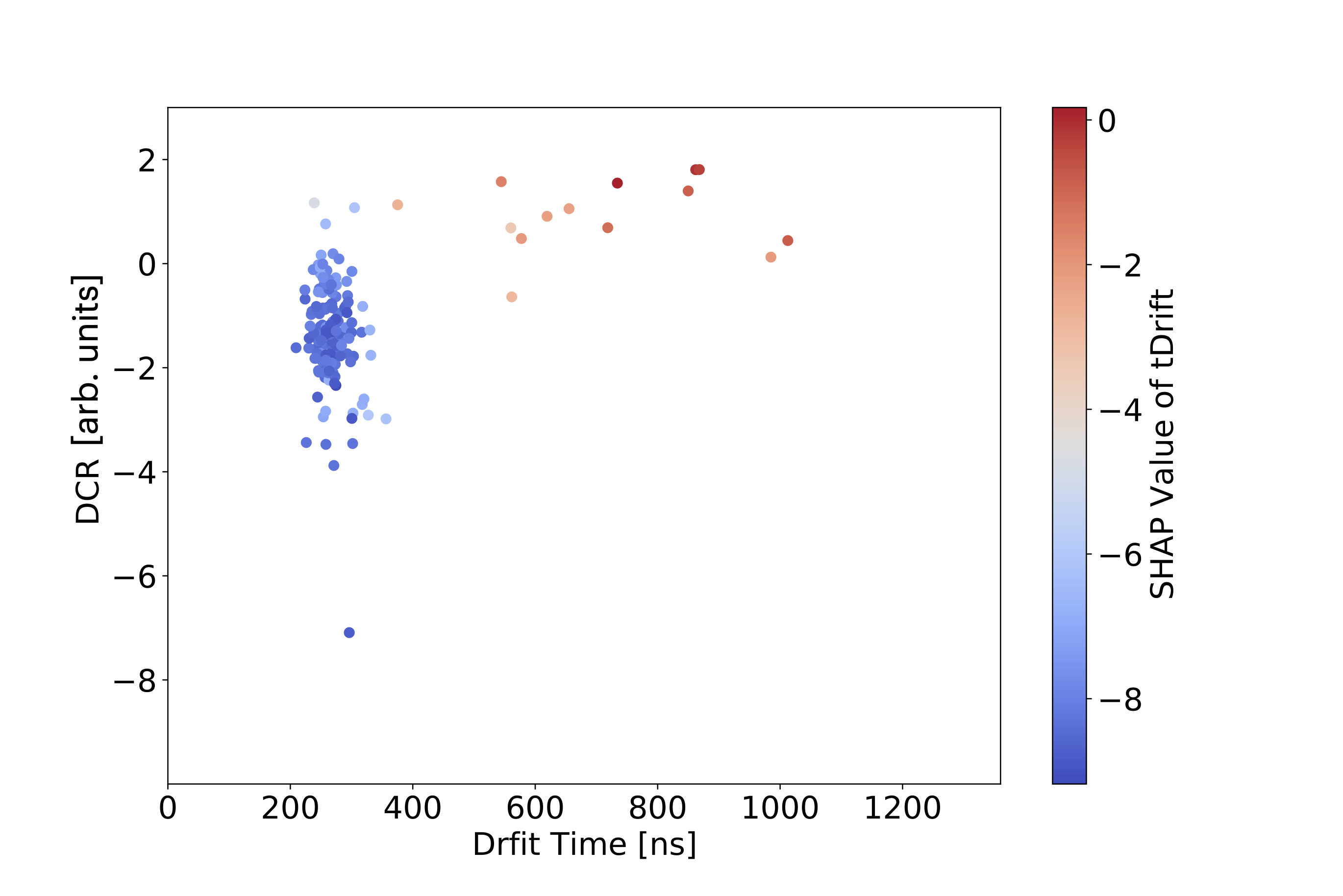}
        \caption{}
        \label{fig:tdrift_dependency_abdt}
    \end{subfigure}
    \caption{(a) Feature importance plot of the \abdt. (b) The scatter plot of raw \texttt{DCR} vs. \texttt{tDrift} for all outperforming genunie \ap~events. The color of each dot represents the sum of Shapley value assigned to both \texttt{tDrift} and \texttt{tDrift50}. Higher magnitude represents more important contributions from drift time.}
   \label{fig:pressure_map}
\end{figure*}

The non-linear dependency happens primarily on events with drift time below 400\,ns. These events happen near the point-contact and drift almost immediately to it. As shown in Figure~\ref{fig:tdrift_dependency}, these fast-drifting events possess excessively high \texttt{AvsE} and will be classified as signals even if the waveform is multi-site. In the standard \textsc{Majorana} analysis, we use the high \texttt{AvsE} cut and the \texttt{LQ} cut to remove these events near the point-contact. In this analysis, the \mbdt~assigns a negative Shapley value according to the drift time to compensate for the higher-than-usual \texttt{AvsE} values. To demonstrate this, we selected a single event from this region (the magenta diamond on Figure~\ref{fig:tdrift_dependency}) and showed its Shapley forces in Figure~\ref{fig:force_plot}. Although the excessively high \texttt{AvsE} produces an overwhelmingly positive force, \mbdt~recognizes the non-linear drift time dependency and assigns negative forces to \texttt{tDrift} and \texttt{tDrift50} to counteract the positive force. The equilibrium position is at 0.34, which falls below the cutting threshold of \mbdt. Therefore, this event is rejected by \mbdt~but accepted by standard \texttt{AvsE}. Without explicit programming, the \mbdt~learns the linear and non-linear correlation from data and handles them correctly to produce better background tagging efficiency.

\subsection{Interpreting $\alpha$BDT}\label{subsec:interp_abdt}
We used a similar approach to interpret the \abdt. Since there are only 3,562 genuine \ap~events, we collected all the genuine \ap~events as backgrounds and 3,562 randomly sampled \th~DEP events as signals to form the interpretability dataset. The summary plot is shown in Figure~\ref{fig:abdt_feature}. As expected, raw \texttt{DCR} is the most important feature in making a classification decision. \texttt{tDrift50} is the second most important feature, indicating that the \abdt~is mainly performing a drift-time correction on raw \texttt{DCR} to enhance its classification power. Similar to the \mbdt, \texttt{detType} is the least important feature since it is redundant with \texttt{channel}.

Outperforming events are collected from the interpretability dataset to understand the additional classification power of the \abdt. Since the signal sacrifice of the \abdt~and \texttt{DCR} is negligible, the outperforming dataset is defined as genuine \ap~events rejected by the \abdt~but accepted by the standard DCR. Figure~\ref{fig:tdrift_dependency_abdt} shows the joint distribution of drift time and standard \texttt{DCR} on a 2D scatter plot on outperforming events. These events form a cluster near a drift time of 200\,ns, indicating that they are surface \ap~events near the point contact. After creation, these events drift to the point contact almost immediately, leaving almost no delayed charge on the passivated surface thus violating the first principle of the \texttt{DCR} cut. However, the fast-drifting nature allows the \abdt~to efficiently tag these events based on their drift-time, thus outperforming the traditional analysis. As \abdt~interpretability study revealed the importance of these backgrounds, a dedicated high \texttt{AvsE} cut is introduced into the standard \textsc{Majorana} analysis to reject them. High \texttt{AvsE} turns out to also reject multisite event near the point contact as we discussed in Section~\ref{subsec:interp_msbdt}.

\begin{figure}[h]
    \centering
    \includegraphics[width=1.0\linewidth,,trim={2pc 2pc 2pc 3pc},clip]{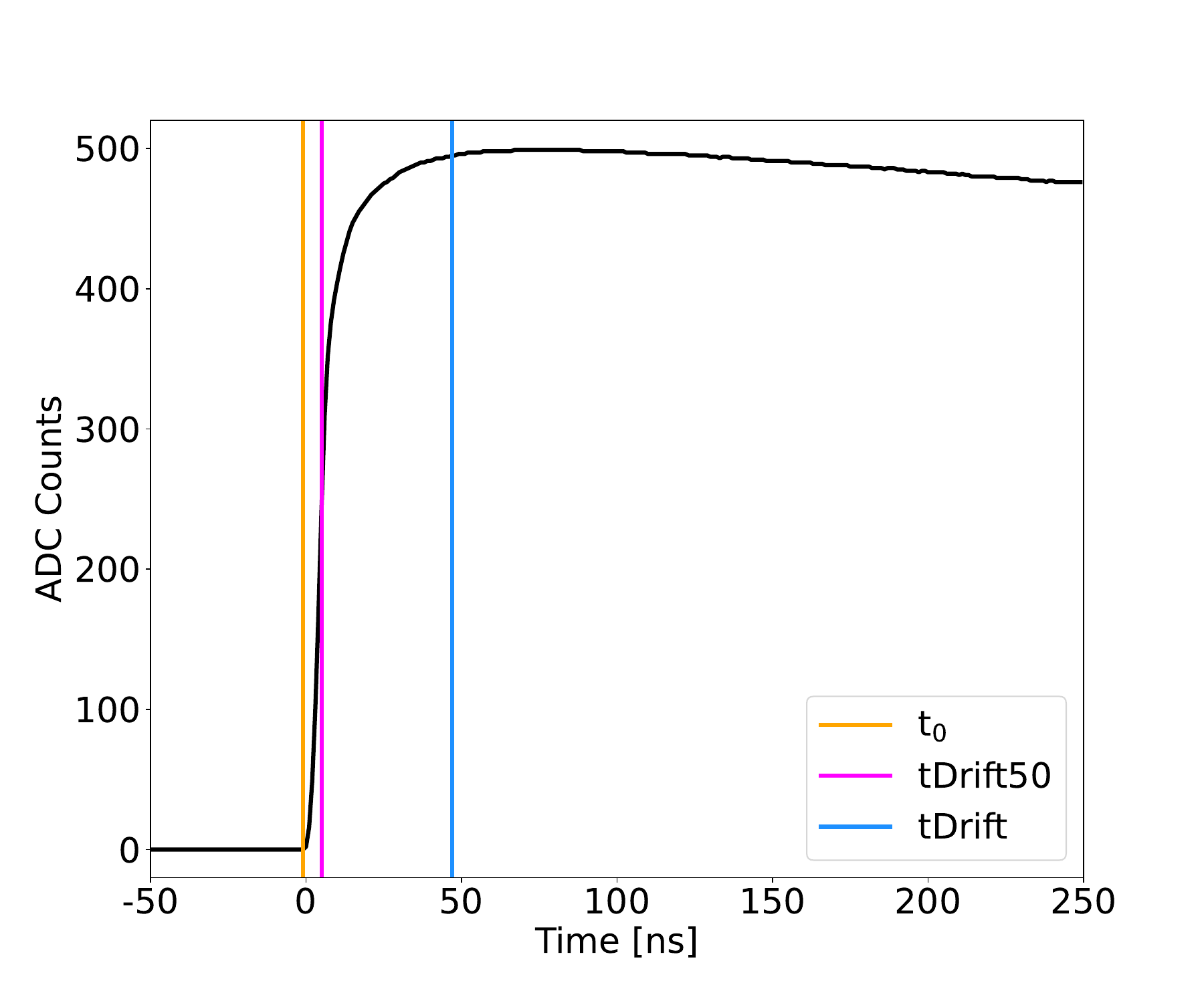}
    \caption{A typical surface \ap~event waveform~(black) near the point-contact of \textsc{Majorana} PPC detector. t$_{0}$ is the start of charge deposition. The time interval of \texttt{tDrift50} and \texttt{tDrift} features are shown.}
    \label{fig:surface_ap_near_pc}
\end{figure}
The interpretability study also shows that \texttt{tDrift50} is more important in the \abdt~model than \texttt{tDrift}. This can be explained by the difference between the calculation of these two parameters. A typical outperforming event is shown in Figure~\ref{fig:surface_ap_near_pc}. When a surface \ap~event happens near the point-contact, the charge deposition starts almost immediately, leading to a sharp rising edge of the waveform. On the other hand, the passivated surface reduces the drift speed of charges comparatively further away from the point-contact. This effect delays the completion of charge deposition, leading to a rounded top of the waveform. Since \texttt{tDrift} is calculated from the the start of the rise to the time when the waveform reaches 99\% of its maximum amplitude, the rounded-top structure significantly increases the value of tDrift, allowing it to appear as a slowly drifting event, thus escape the low drift-time/high \texttt{AvsE} cut. However, \texttt{tDrift50} is immune to the rounded-top structure since it is caluclated only up to 50\% of the waveform amplitude. The interpretability study suggests that a \texttt{tDrift50}-based cut could be developed to further benefit the \ap~rejection in future first-principle analyses.

The interpretability study allows machine learning analysis to unveil physics in germanium detectors. Leveraging multivariate correlations and automatic categorization, the BDT was able to outperform individual PSD parameters and match both the GAT and the ORNL analyses, as discussed in Section~\ref{sec:result} with less detector-by-detector calibration. Furthermore, the interpretability study leverages the additional classification power to reveal the importance of new background categories. This eventually led to the implementation of a high \texttt{AvsE} cut in the standard \textsc{Majorana} analysis and suggests a new direction for future improvement. The reciprocal relationship between the machine learning analysis and the traditional, first-principle analysis revealed by the interpretability study, demonstrates that an interpretable machine learning analysis can not only outperform but also benefit the traditional analysis.

\section{conclusion}\label{sec:conclusion}
In this work, we have presented the first machine learning analysis for the \mjd; this is also the first interpretable machine learning analysis of any germanium detector experiment. This analysis contains two parts: learning from the data to improve background rejections and learning from the machine to understand classification power. Leveraging gradient boosted decision trees and data augmentation, this analysis outperforms the the individual PSD parameters and match the overall results of the highly optimized standard \textsc{Majorana} analysis~\cite{Majorana_prl}. Learning from data also closes the gap between two independently developed analyses applied to different types of detectors.

For the first time in the field, a thorough machine interpretability study is conducted, leveraging the Shapley value in coalitional game theory. This study not only justifies BDT's capability to capture multivariate correlations but also to independently discover new background categories to reveal its importance. The machine learning analysis and the standard \textsc{Majorana} analysis established a reciprocal relationship through the interpretability study. Since BDT model is widely used in the particle and nuclear physics community~\cite{MicroBooNE_bdt,CERN_bdt,BDT_Atlas,BDT_atlas2,BDT_cms,BDT_miniboone,BDT_tq}, this work provides a template for interpreting the BDT model to gain more physical insight and even make new scientific discovery.

This work has focused on developing and interpreting the first machine learning analysis for \mjd. The data-driven nature of this analysis allows a straightforward generalization to different germanium detector experiments, especially the next-generation tonne-scale experiment LEGEND-1000~\cite{legend_pcdr}. Given the large number of detectors, detector- and run-level tuning may be time-consuming in LEGEND. In that case, the BDT's ability to simultaneously train on all detectors would be highly beneficial. Furthermore, the interpretability study allows us to unravel the black-box nature of machine learning models to reveal underlying physics and independently discover new background categories without explicit programming. We intend to apply this model to LEGEND data, which could enable improvements in background rejection, and possibly help us gain a more nuanced understanding of the detector performance. Our future work involves using more powerful and versatile machine learning models such as recurrent neural network~(RNN). RNN can be trained directly on the \textsc{Demonstrator}'s waveform, which opens up an entirely new avenue for more machine learning applications.


\begin{acknowledgments}
This material is based upon work supported by the U.S.~Department of Energy, Office of Science, Office of Nuclear Physics under contract / award numbers DE-AC02-05CH11231, DE-AC05-00OR22725, DE-AC05-76RL0130, DE-FG02-97ER41020, DE-FG02-97ER41033, DE-FG02-97ER41041, DE-SC0012612, DE-SC0014445, DE-SC0018060, DE-SC0022339, and LANLEM77/LANLEM78. We acknowledge support from the Particle Astrophysics Program and Nuclear Physics Program of the National Science Foundation through grant numbers MRI-0923142, PHY-1003399, PHY-1102292, PHY-1206314, PHY-1614611, PHY-1812409, PHY-1812356, PHY-2111140, and PHY-2209530. We gratefully acknowledge the support of the Laboratory Directed Research \& Development (LDRD) program at Lawrence Berkeley National Laboratory for this work. We gratefully acknowledge the support of the U.S.~Department of Energy through the Los Alamos National Laboratory LDRD Program and through the Pacific Northwest National Laboratory LDRD Program for this work.  We gratefully acknowledge the support of the South Dakota Board of Regents Competitive Research Grant. We acknowledge the support of the Natural Sciences and Engineering Research Council of Canada, funding reference number SAPIN-2017-00023, and from the Canada Foundation for Innovation John R.~Evans Leaders Fund.  This research used resources provided by the Oak Ridge Leadership Computing Facility at Oak Ridge National Laboratory and by the National Energy Research Scientific Computing Center~(NERSC), a U.S.~Department of Energy Office of Science User Facility at Lawrence Berkeley National Laboratory. We thank our hosts and colleagues at the Sanford Underground Research Facility for their support.
\end{acknowledgments}

\nocite{*}

\bibliography{BDT_paper}

\end{document}

%% file: authors.tex
\newcommand{\ITEP}{National Research Center ``Kurchatov Institute'' Institute for Theoretical and Experimental Physics, Moscow, 117218 Russia}
\newcommand{\JINR}{Joint Institute for Nuclear Research, Dubna, 141980 Russia} 
\newcommand{\lbnl}{Nuclear Science Division, Lawrence Berkeley National Laboratory, Berkeley, CA 94720, USA}
\newcommand{\lbnle}{Engineering Division, Lawrence Berkeley National Laboratory, Berkeley, CA 94720, USA}
\newcommand{\lanl}{Los Alamos National Laboratory, Los Alamos, NM 87545, USA}
\newcommand{\queens}{Department of Physics, Engineering Physics and Astronomy, Queen's University, Kingston, ON K7L 3N6, Canada}
\newcommand{\uw}{Center for Experimental Nuclear Physics and Astrophysics, and Department of Physics, University of Washington, Seattle, WA 98195, USA}
\newcommand{\unc}{Department of Physics and Astronomy, University of North Carolina, Chapel Hill, NC 27514, USA}
\newcommand{\duke}{Department of Physics, Duke University, Durham, NC 27708, USA}
\newcommand{\ncsu}{Department of Physics, North Carolina State University, Raleigh, NC 27695, USA}	
\newcommand{\ornl}{Oak Ridge National Laboratory, Oak Ridge, TN 37830, USA}
\newcommand{\ou}{Research Center for Nuclear Physics, Osaka University, Ibaraki, Osaka 567-0047, Japan}
\newcommand{\pnnl}{Pacific Northwest National Laboratory, Richland, WA 99354, USA}
\newcommand{\ttu}{Tennessee Tech University, Cookeville, TN 38505, USA}
\newcommand{\sdsmt}{South Dakota Mines, Rapid City, SD 57701, USA}
\newcommand{\usc}{Department of Physics and Astronomy, University of South Carolina, Columbia, SC 29208, USA}
\newcommand{\usd}{Department of Physics, University of South Dakota, Vermillion, SD 57069, USA}  
\newcommand{\ut}{Department of Physics and Astronomy, University of Tennessee, Knoxville, TN 37916, USA}
\newcommand{\tunl}{Triangle Universities Nuclear Laboratory, Durham, NC 27708, USA}
\newcommand{\mpi}{Max-Planck-Institut f\"{u}r Physik, M\"{u}nchen, 80805, Germany}
\newcommand{\tum}{Physik Department and Excellence Cluster Universe, Technische Universit\"{a}t, M\"{u}nchen, 85748 Germany}
\newcommand{\williams}{Physics Department, Williams College, Williamstown, MA 01267, USA}
\newcommand{\ciemat}{Centro de Investigaciones Energ\'{e}ticas, Medioambientales y Tecnol\'{o}gicas, CIEMAT 28040, Madrid, Spain}
\newcommand{\iu}{Department of Physics, Indiana University, Bloomington, IN 47405, USA}
\newcommand{\iuceem}{IU Center for Exploration of Energy and Matter, Bloomington, IN 47408, USA}

\author{I.J.~Arnquist}\affiliation{\pnnl} 
\author{F.T.~Avignone~III}\affiliation{\usc}\affiliation{\ornl}
\author{A.S.~Barabash}\affiliation{\ITEP}
\author{C.J.~Barton}\affiliation{\usd}	
\author{K.H.~Bhimani}\affiliation{\unc}\affiliation{\tunl} 
\author{E.~Blalock}\affiliation{\ncsu}\affiliation{\tunl} 
\author{B.~Bos}\affiliation{\unc}\affiliation{\tunl} 
\author{M.~Busch}\affiliation{\duke}\affiliation{\tunl}	
\author{M.~Buuck}\altaffiliation{Present address: SLAC National Accelerator Laboratory, Menlo Park, CA 94025, USA}\affiliation{\uw} 
\author{T.S.~Caldwell}\affiliation{\unc}\affiliation{\tunl}	
\author{Y-D.~Chan}\affiliation{\lbnl}
\author{C.D.~Christofferson}\affiliation{\sdsmt} 
\author{P.-H.~Chu}\affiliation{\lanl} 
\author{M.L.~Clark}\affiliation{\unc}\affiliation{\tunl} 
\author{C.~Cuesta}\affiliation{\ciemat}	
\author{J.A.~Detwiler}\affiliation{\uw}	
\author{Yu.~Efremenko}\affiliation{\ut}\affiliation{\ornl}
\author{S.R.~Elliott}\affiliation{\lanl}
\author{G.K.~Giovanetti}\affiliation{\williams}  
\author{M.P.~Green}\affiliation{\ncsu}\affiliation{\tunl}\affiliation{\ornl}   
\author{J.~Gruszko}\affiliation{\unc}\affiliation{\tunl} 
\author{I.S.~Guinn}\affiliation{\unc}\affiliation{\tunl} 
\author{V.E.~Guiseppe}\affiliation{\ornl}	
\author{C.R.~Haufe}\affiliation{\unc}\affiliation{\tunl}	
\author{R.~Henning}\affiliation{\unc}\affiliation{\tunl}
\author{D.~Hervas~Aguilar}\affiliation{\unc}\affiliation{\tunl} 
\author{E.W.~Hoppe}\affiliation{\pnnl}
\author{A.~Hostiuc}\affiliation{\uw} 
\author{M.F.~Kidd}\affiliation{\ttu}	
\author{I.~Kim}\affiliation{\lanl} 
\author{R.T.~Kouzes}\affiliation{\pnnl}
\author{T.E.~Lannen~V}\affiliation{\usc} 
\author{A.~Li}\email{Corresponding Author. Email: liaobo77@ad.unc.edu}\affiliation{\unc}\affiliation{\tunl} 
\author{J.M. L\'opez-Casta\~no}\affiliation{\ornl} 
\author{E.L.~Martin}\altaffiliation{Present address: Duke University, Durham, NC 27708}\affiliation{\unc}\affiliation{\tunl}	
\author{R.D.~Martin}\affiliation{\queens}	
\author{R.~Massarczyk}\affiliation{\lanl}		
\author{S.J.~Meijer}\affiliation{\lanl}	
\author{T.K.~Oli}\affiliation{\usd}  
\author{G.~Othman}\altaffiliation{Present address: Universit{\"a}t Hamburg, Institut f{\"u}r Experimentalphysik, Hamburg, Germany}\affiliation{\unc}\affiliation{\tunl} 
\author{L.S.~Paudel}\affiliation{\usd} 
\author{W.~Pettus}\affiliation{\iu}\affiliation{\iuceem}	
\author{A.W.P.~Poon}\affiliation{\lbnl}
\author{D.C.~Radford}\affiliation{\ornl}
\author{A.L.~Reine}\affiliation{\unc}\affiliation{\tunl}	
\author{K.~Rielage}\affiliation{\lanl}
\author{N.W.~Ruof}\affiliation{\uw}	
\author{D.C.~Schaper}\affiliation{\lanl} 
\author{D.~Tedeschi}\affiliation{\usc}		
\author{R.L.~Varner}\affiliation{\ornl}  
\author{S.~Vasilyev}\affiliation{\JINR}	
\author{J.F.~Wilkerson}\affiliation{\unc}\affiliation{\tunl}\affiliation{\ornl}    
\author{C.~Wiseman}\affiliation{\uw}		
\author{W.~Xu}\affiliation{\usd} 
\author{C.-H.~Yu}\affiliation{\ornl}

\collaboration{{\sc{Majorana}} Collaboration}
\noaffiliation